\pdfoutput=1
\newcommand*{\ATLASLATEXPATH}{latex/}
\documentclass[UKenglish,cernpreprint,texlive=2016]{\ATLASLATEXPATH atlasdoc}
\usepackage[subfigure]{\ATLASLATEXPATH atlaspackage}
\usepackage{\ATLASLATEXPATH atlasbiblatex}
\usepackage{multirow}
\usepackage{enumerate}
\usepackage [BSM, other,process,particle]{\ATLASLATEXPATH atlasphysics}

\graphicspath{{logos/}}

\usepackage{HPhiGamma-defs}

\AtlasTitle{Search for exclusive Higgs and $Z$ boson decays to $\phi\gamma$ and $\rho\gamma$ with the ATLAS detector}

\author{The ATLAS Collaboration}

\date{\today}

\PreprintIdNumber{CERN-EP-2017-273}

\AtlasJournal{JHEP}
\AtlasJournalRef{JHEP 07 (2018) 127}
\AtlasDOI{DOI:10.1007/JHEP07(2018)127}

\AtlasAbstract{%
A search for the exclusive decays of the Higgs and $Z$ bosons to
a $\phi$ or $\rho$ meson and a photon is performed with a $pp$ collision
data sample corresponding to an integrated luminosity of up to
$35.6\,\ifb$ collected at $\sqrt{s}=13\,\TeV$ with the ATLAS detector at
the CERN Large Hadron Collider. 
These decays have been suggested as a probe of the Higgs boson
couplings to light quarks.
No significant excess of events is observed above the background, as expected from the Standard Model. Upper limits at 95\% confidence
level were obtained on the branching fractions of the Higgs boson
decays to $\phi\gamma$ and $\rho\gamma$ of $5.0\times10^{-4}$ and
$10.4\times10^{-4}$, respectively.
The corresponding 95\%
confidence level upper limits for the $Z$ boson decays are
$0.7\times10^{-6}$ and $4.0\times10^{-6}$ for $\phi\gamma$ and
$\rho\gamma$, respectively.  
}

\hypersetup{pdftitle={ATLAS document},pdfauthor={The ATLAS Collaboration}}

\def\ee{\ensuremath{e^+ e^-}}%
\def\epem{\ensuremath{e^+ e^-}}%
\def\mumu{\ensuremath{\mathrm{\mu^+ \mu^-}}}%
\def\tautau{\ensuremath{\mathrm{\tau^+ \tau^-}}}%

\def\leplep{\ensuremath{\ell^+ \ell^-}}%
\def\lnu{\ensuremath{\ell \nu}}%

\def\ptmuq#1 {\ensuremath{p_{\textrm{T}}^{\,\mu^{#1}}}}

\def\etamuq#1 {\ensuremath{\eta^{\,\mu^{#1}}}}

\def\ptgamma {\ensuremath{p_{\textrm{T}}^{\gamma}}}
\def\etagamma {\ensuremath{\eta^{\gamma}}}

\def\phiKK{\ensuremath{\phi\to K^{+}K^{-}}}
\def\KK{\ensuremath{K^{+}K^{-}}}

\newcommand{\tabscript}[3]{%
  \setlength{\fboxrule}{0pt}%
  \fbox{\ensuremath{#1^{#2}_{#3}}}%

  }
\begin{document}

\maketitle

\section{Introduction}
\label{sec:intro}

Following the observation~\cite{HIGG-2012-27,CMS-HIG-12-028} of a
Higgs boson, $H$, with a mass of approximately
125~GeV~\cite{CMS-HIG-14-042} by the ATLAS and CMS collaborations at the Large Hadron Collider (LHC), the
properties of its interactions with the electroweak gauge bosons have
been measured
extensively~\cite{CMS-HIG-15-002,HIGG-2014-06,CMS-HIG-14-009}.
The coupling of the Higgs boson to leptons has been established
through the observation of the $H\to\tau^+\tau^-$
channel~\cite{CMS-HIG-15-002,HIGG-2013-32,CMS-HIG-13-004},
while in the quark sector indirect evidence is available for the
coupling of the Higgs boson to the
top-quark~\cite{CMS-HIG-15-002} and evidence for the Higgs
boson decays into $b\bar{b}$ has been recently presented~\cite{HIGG-2016-29,CMS-HIG-16-044}.
Despite this progress, the Higgs
boson interaction with the fermions of the first and second generations
is still to be confirmed experimentally.
In the Standard Model (SM), Higgs boson interactions to fermions are implemented through Yukawa
couplings, while a wealth of beyond-the-SM theories predict
substantial modifications. Such scenarios include the Minimal Flavour Violation
framework~\cite{DAmbrosio:2002ex}, the Froggatt--Nielsen
mechanism~\cite{Froggatt:1978nt}, the Higgs-dependent Yukawa couplings
model~\cite{Giudice:2008uua}, the Randall--Sundrum family of models
\cite{Randall:1999ee}, and the possibility of the Higgs boson being a
composite pseudo-Goldstone boson~\cite{Dugan:1984hq}. An overview of
relevant models of new physics is provided in
Ref.~\cite{deFlorian:2016spz}.

The rare decays of the Higgs boson into a heavy quarkonium state,
$\Jpsi$ or $\Upsilon(nS)$ with $n=1,2,3$, and a photon have been suggested for probing the charm- and bottom-quark couplings to the Higgs
boson~\cite{VatoHBranching,HBranching,Bodwin:2014bpa,Bodwin:2016edd}
and have already been searched for by the ATLAS
Collaboration~\cite{HIGG-2014-03}, resulting in 95\% confidence level (CL) upper limits of
$1.5\times 10^{-3}$ and $\left(1.3,1.9,1.3\right)\times 10^{-3}$ on the branching fractions,
respectively. The $H\to\Jpsi\gamma$ decay mode has also been searched for by the
CMS Collaboration~\cite{CMS-HIG-14-003}, yielding the same upper
limit. The corresponding SM predictions for these branching fractions~\cite{Koenig:2015pha} are
${\cal B}\left(H\to\Jpsi\gamma\right)=\left(2.95\pm 0.17\right)\times 10^{-6}$ and
${\cal B}\left(H\to\Upsilon(nS)\gamma\right)=\left(4.6^{+1.7}_{-1.2},2.3^{+0.8}_{-1.0},2.1^{+0.8}_{-1.1}\right)\times
10^{-9}$. 
The prospects for observing and studying exclusive Higgs boson decays into a
meson and a photon with an upgraded High Luminosity
LHC~\cite{deFlorian:2016spz} or a future hadron
collider~\cite{Contino:2016spe} have also been studied.

Currently, the light ($u,d,s$) quark couplings to the Higgs boson are
loosely constrained by existing data on the total Higgs boson width, while the large multijet background
at the LHC inhibits the study of such
couplings with inclusive $H\to q\bar{q}$ decays.
Rare exclusive decays of the Higgs boson into a light meson, $M$, and a
photon, $\gamma$, have been suggested as a probe of the couplings of
the Higgs boson to
light quarks and would allow a search for
potential deviations from the SM prediction~\cite{Kagan:2014ila,Koenig:2015pha,Perez:2015lra}.
Specifically, the observation of the Higgs boson decay to a $\phi$ or $\rho(770)$ (denoted as $\rho$ in the following) meson and a photon
would provide sensitivity to its couplings to the strange-quark, and the up-
and down-quarks, respectively. The expected SM branching
fractions are
${\cal B}\left(H\to\phi\gamma\right)=(2.31\pm0.11)\times 10^{-6}$
and
${\cal B}\left(H\to\rho\gamma\right)=(1.68\pm0.08)\times10^{-5}$%
~\cite{Koenig:2015pha}. 
The decay amplitude receives two main contributions that interfere
destructively. The first is referred to as ``direct'' and proceeds through
the $H\to q\bar{q}$ coupling, where subsequently a photon is emitted
before the $q\bar{q}$ hadronises exclusively to $M$. 
The second is referred to as ``indirect'' and proceeds via the
$H\to\gamma\gamma$ coupling followed by the fragmentation $\gamma^*\to
M$. In the SM, owing to the smallness of the light-quark Yukawa
couplings, the latter amplitude dominates, despite being loop
induced. As a result, the expected branching fraction predominantly
arises from the ``indirect'' process, while the Higgs boson couplings to
the light quarks are probed by searching for modifications of this
branching fraction due to changes in the ``direct'' amplitude.

This paper describes a search for Higgs boson decays into the
exclusive final states $\phi\gamma$ and $\rho\gamma$. The decay
$\phi\to K^{+}K^{-}$ is used to reconstruct the $\phi$ meson, and 
the decay $\rho\to\pi^+\pi^-$ is used to reconstruct the $\rho$
meson. The branching fractions of the respective meson decays are well known and are accounted for when
calculating the expected signal yields. The presented search uses approximately 13
times more
integrated luminosity than the first search for
$H\to\phi\gamma$ decays~\cite{HIGG-2016-05}, which led to a 95\%
CL upper limit of ${\cal
  B}\left(H\to\phi\gamma\right)<1.4\times 10^{-3}$, assuming SM
production rates of the Higgs boson. Currently, no other experimental information about the
$H\to\rho\gamma$ decay mode exists.

The searches for the analogous decays of the $Z$ boson into a meson and
a photon are also presented in this paper. These have been
theoretically studied~\cite{Huang:2014cxa,Grossmann:2015lea} as 
a unique precision test of the SM and the factorisation approach in
quantum chromodynamics (QCD), in an environment where the power corrections
in terms of the QCD energy scale over the vector boson's mass are
small~\cite{Grossmann:2015lea}.
The large $Z$ boson production cross section at the LHC means that rare
$Z$ boson decays can be probed at branching fractions much smaller
than for Higgs boson decays into the same final states. The SM branching fraction
predictions for the decays considered in this paper are ${\cal
B}\left(Z\to\phi\gamma\right)=(1.04\pm0.12)\times
10^{-8}$~\cite{Grossmann:2015lea,Huang:2014cxa} and ${\cal
B}\left(Z\to\rho\gamma\right)=(4.19\pm0.47)\times
10^{-8}$~\cite{Grossmann:2015lea}.
The first search for $Z\to\phi\gamma$ decays by the ATLAS Collaboration was presented in
Ref.~\cite{HIGG-2016-05} and a 95\% CL upper limit of
${\cal B}\left(Z\to\phi\gamma\right)<8.3\times 10^{-6}$ was
obtained. So far no direct experimental information about the decay
$Z\to\rho\gamma$ exists.

\section{ATLAS detector}
\label{sec:atlas}

ATLAS~\cite{PERF-2007-01} is a multi-purpose particle physics detector
with a forward-backward symmetric cylindrical geometry and near 4$\pi$
coverage in solid angle.\footnote{ATLAS uses a right-handed coordinate
  system with its origin at the nominal interaction point (IP) in the
  centre of the detector and the $z$-axis along the beam pipe. The
  $x$-axis points from the IP to the centre of the LHC ring, and the
  $y$-axis points upward. Cylindrical coordinates $(r,\phi)$ are used
  in the transverse plane, $\phi$ being the azimuthal angle around the
  $z$-axis. The pseudorapidity is defined in terms of the polar angle
  $\theta$ as $\eta=-\ln\tan(\theta/2)$.} It consists of an inner
tracking detector surrounded by a thin superconducting solenoid,
electromagnetic and hadronic calorimeters, and a muon spectrometer.

The inner tracking detector (ID) covers the pseudorapidity range
$|\eta|<2.5$, and is surrounded by a thin superconducting solenoid
providing a 2~T magnetic field. At small radii, a high-granularity
silicon pixel detector covers the vertex region and typically provides
three measurements per track. A new innermost pixel-detector layer,
the insertable B-layer, was added before $13\,\TeV$\ data-taking began in 2015
and provides an additional measurement at a radius of about
\SI{33}{\milli\meter} around a new and thinner beam
pipe~\cite{ATLAS-TDR-19}. The pixel detectors are followed by a
silicon microstrip tracker, which typically provides four space-point measurements
per track. The silicon detectors are complemented by a gas-filled
straw-tube transition radiation tracker, which enables radially
extended track reconstruction up to $|\eta| = 2.0$, with typically 35
measurements per track.

The calorimeter system covers the pseudorapidity range $|\eta| < 4.9$.
A high-granularity lead/liquid-argon (LAr) sampling electromagnetic
calorimeter covers the region $|\eta|<3.2$, with an additional thin
LAr presampler covering $|\eta| < 1.8$ to correct for 
energy losses upstream. The electromagnetic calorimeter is divided
into a barrel section covering $|\eta| < 1.475$ and two endcap
sections covering $1.375 < |\eta| < 3.2$. For $|\eta| < 2.5$ it is
divided into three layers in depth, which are finely segmented in
$\eta$ and $\phi$.  A steel/scintillator-tile calorimeter provides
hadronic calorimetry in the range $|\eta|<1.7$. LAr technology, with
copper as absorber, is used for the hadronic calorimeters in the
endcap region, $1.5<|\eta|<3.2$.  The solid-angle coverage is
completed with forward copper/LAr and tungsten/LAr calorimeter modules
in $3.1 < |\eta| < 4.9$, optimised for electromagnetic and hadronic
measurements, respectively.

The muon spectrometer surrounds the calorimeters and comprises
separate trigger and high-precision tracking chambers measuring the
deflection of muons in a magnetic field provided by three air-core
superconducting toroids. 

A two-level trigger and data acquisition system is used to provide an online selection and record
events for offline analysis~\cite{TRIG-2016-01}. The level-1
trigger is implemented in hardware and uses a subset of detector
information to reduce the event rate to \SI{100}{\kilo\hertz} or less from the maximum LHC collision rate of \SI{40}{\mega\hertz}. It is
followed by a software-based high-level trigger which filters events using the
full detector information and records events for detailed offline analysis
at an average rate of \SI{1}{\kilo\hertz}.

\section{Data and Monte Carlo simulation}
\label{sec:datamc}

The search is performed with a sample of $pp$ collision data recorded at a
centre-of-mass energy $\rts=13\,\TeV$. Events are retained for further analysis
only if they were collected under stable LHC beam conditions and the detector
was operating normally. This results in an integrated luminosity of 35.6 and
32.3~\ifb\ for the $\phi\gamma$ and $\rho\gamma$ final states, respectively.
The integrated luminosity of the data sample has an uncertainty of $3.4\%$
derived using the method described in Ref.~\cite{DAPR-2013-01}. 

The $\phi\gamma$ and $\rho\gamma$ data samples used in this analysis were each collected with a specifically designed trigger.
Both triggers require an isolated photon with a transverse momentum, $\pt$, greater
than $35\,\GeV$ and an isolated pair of ID tracks, one of which must
have a $\pt$ greater than $15\,\GeV$, associated with a topological
cluster of calorimeter cells~\cite{PERF-2014-07} with a transverse energy
greater than $25\,\GeV$.  The photon part of the trigger follows the same process
as the inclusive photon trigger requiring an electromagnetic cluster in the
calorimeter consistent with a photon and is described with more detail in
Ref.~\cite{TRIG-2016-01}, while
requirements on the ID tracks are applied in the high-level trigger
through an appropriately modified version of the $\tau$-lepton trigger
algorithms which are described in more detail in Ref.~\cite{ATLAS-CONF-2017-061}.
The trigger for the $\phi\gamma$ final state was introduced in
September 2015. This trigger requires that the invariant mass of the
pair of tracks, under the charged-kaon hypothesis, is in the range 987--1060\,$\MeV$,
consistent with the $\phi$ meson mass. The trigger
efficiency for both the Higgs and
$Z$ boson signals is approximately 83\% with respect to the offline
selection, as described in Section~\ref{sec:selection}.
The corresponding trigger for the $\rho\gamma$ final state was introduced
in May 2016. This trigger requires the invariant mass of the pair of tracks,
under the charged-pion hypothesis, to be in the range 475--1075\,$\MeV$
to include the bulk of the broad $\rho$ meson mass distribution. The trigger efficiency
for the Higgs boson signal is approximately 78\% and for the $Z$ boson signal is approximately 72\% with
respect to the offline selection.

Higgs~boson production through the gluon--gluon fusion ($ggH$) and vector-boson fusion
(VBF) processes was modelled up to next-to-leading order (NLO) in $\alphas$
using the \POWHEGBOX v2 Monte Carlo (MC) event
generator~\cite{powheg1,powheg2,powheg3,powheg4,powheg5} with CT10 parton
distribution functions~\cite{Lai:2010vv}.
\POWHEGBOX was interfaced with the
\PYTHIAV{8.186} MC event
generator~\cite{Sjostrand:2007gs,Sjostrand:2006za} to model the parton shower,
 hadronisation and underlying event. The corresponding parameter values were set according to
the AZNLO tune~\cite{STDM-2012-23}.
Additional contributions from the associated production of a Higgs
boson and a $W$ or $Z$ boson (denoted by $WH$ and $ZH$, respectively) are
modelled by the \PYTHIAV{8.186} MC event
generator
with NNPDF23LO
parton distribution functions~\cite{Ball:2012cx} and the A14 tune for hadronisation and the underlying event~\cite{ATL-PHYS-PUB-2014-021}. The production
rates and kinematic distributions for the SM Higgs boson with $\mH =125\,\gev$ are assumed throughout. These were obtained from Ref.~\cite{deFlorian:2016spz} and are summarised below.  The $ggH$
production rate is normalised such that it reproduces the total cross
section predicted by a next-to-next-to-next-to-leading-order
QCD calculation with NLO electroweak corrections
applied~\cite{Anastasiou:2015ema,Anastasiou:2016cez,Actis:2008ug,Anastasiou:2008tj}. The
VBF production rate is normalised to an approximate NNLO QCD cross section
with NLO electroweak corrections
applied~\cite{Ciccolini:2007jr,Ciccolini:2007ec,Bolzoni:2010xr}. The
$WH$ and $ZH$ production rates are normalised to cross sections
calculated at next-to-next-to-leading order (NNLO) in QCD with NLO electroweak
corrections~\cite{Brein:2003wg,Denner:2011id} including the NLO QCD
corrections~\cite{Altenkamp:2012sx} for $gg\to ZH$. The expected
signal yield is corrected to include the 2\% contribution from the
production of a Higgs boson in association with a $t\bar{t}$ or a
$b\bar{b}$ pair.

The \POWHEGBOX v2 MC event generator with CT10 parton
distribution functions was also used to model
inclusive $Z$ boson production.
\PYTHIAV{8.186} with
CTEQ6L1 parton distribution functions~\cite{Pumplin:2002vw} and the
AZNLO parameter tune was used to simulate parton showering and
hadronisation.
The prediction is normalised to the total cross section obtained from
the measurement in Ref.~\cite{STDM-2015-03}, which has an uncertainty
of 2.9\%.
The Higgs and $Z$ boson decays were simulated as a cascade of two-body
decays, respecting angular momentum conservation. The meson line shapes were simulated by 
\PYTHIA. The branching fraction for the decay
$\phi\to K^{+}K^{-}$ is $(48.9\pm0.5)\%$ whereas the decay
$\rho\to\pi^+\pi^-$ has a branching fraction close to
$100\%$~\cite{PDG}.
The simulated events were passed through the detailed \GEANT 4 simulation of the
ATLAS detector~\cite{GEANT4,SOFT-2010-01} and processed with the same software used
to reconstruct the data. Simulated pile-up events (additional $pp$ collisions
in the same or nearby bunch crossings) are also included and the distribution of
these is matched to the conditions observed in the data.

\section{Event selection for $\phi\gamma\to K^{+}K^{-}\gamma$ and $\rho\gamma\to\pi^+\pi^-\gamma$ final states}
\label{sec:selection}

The $\phi\gamma$ and $\rho\gamma$ exclusive final states are very
similar. Both final states consist of a pair of oppositely charged
reconstructed ID tracks. The difference is that for the former the
mass of the pair, under the charged-kaon hypothesis for the two
tracks, is consistent with the $\phi$ meson mass, while for the later,
under the charged-pion hypothesis for the tracks, it is consistent
with the $\rho$ meson mass.
Events with a $pp$ interaction vertex reconstructed from at least two ID
tracks with $\pT > 400\,\mev$ are considered in the analysis. Within an event, the primary
vertex is defined as the reconstructed vertex with the largest
$\sum\pT^2$ of associated ID tracks.

Photons are reconstructed from clusters of energy in the
electromagnetic calorimeter. Clusters without matching ID tracks are
classified as unconverted photon candidates while clusters matched to
ID tracks consistent with the hypothesis of a photon conversion into
$\ee$ are classified as converted photon
candidates~\cite{PERF-2013-04}.  Reconstructed photon candidates are
required to have $\ptgamma>35\,\GeV$, $|\etagamma|<2.37$, excluding the
barrel/endcap calorimeter transition region $1.37<|\etagamma|<1.52$,
and to satisfy ``tight'' photon identification
criteria~\cite{PERF-2013-04}. An isolation requirement is imposed to
further suppress contamination from jets. The sum of the transverse
momenta of all tracks within $\Delta R = \sqrt{(\Delta\phi)^{2} +
  (\Delta\eta)^{2}} = 0.2$ of the photon direction, excluding those
associated with the reconstructed photon, is required to be less than
$5\%$ of $\ptgamma$. Moreover, the sum of the transverse momenta of
all calorimeter energy deposits within $\Delta R =
0.4$ of the photon direction, excluding those associated with the
reconstructed photon, is required to be less than $2.45\,\GeV +
0.022\times\ptgamma$. To mitigate the effects of multiple $pp$
interactions in the same or neighbouring bunch crossings, only ID
tracks which originate from the primary vertex are considered in the
photon track-based isolation. For the calorimeter-based isolation the
effects of the underlying event and multiple $pp$ interactions are also
accounted for on an event by event basis using an average underlying event
energy density determined from data, as described in Ref.~\cite{PERF-2013-04}.

Charged particles satisfying the requirements detailed below are
assumed to be a $K^{\pm}$ meson in the $\phi\gamma$ analysis and a
$\pi^\pm$ meson in the $\rho\gamma$ analysis. No further particle
identification requirements are applied. In the following,
when referring to charged particles collectively the term
``charged-hadron candidates'' is used, while when referring to the
charged particles relevant to the $\phi\gamma$ and the
$\rho\gamma$ analyses the terms ``kaon candidates'' and ``pion candidates'' are
used, respectively, along with the corresponding masses. A pair of oppositely-charged charged-hadron candidates is referred to collectively as $M$.

Charged-hadron candidates are reconstructed from ID tracks which are
required to have $|\eta|<2.5$, $\pt>15\,\GeV$ and to satisfy basic
quality criteria, including a requirement on the number of hits in the
silicon detectors~\cite{ATL-PHYS-PUB-2015-051}.
The $\phiKK$ and $\rho\to\PiPi$ decays are reconstructed from pairs of
oppositely charged-hadron candidates; the candidate with the
higher $\pt$, referred to as the leading charged-hadron candidate, is required to have $\pt>20\,\GeV$.

Pairs of charged-hadron candidates are selected based on their
invariant masses. Those with an invariant mass, under the charged-kaon
hypothesis, $m_{K^{+}K^{-}}$ between $1012\,\MeV$ and $1028\,\MeV$ are
selected as $\phiKK$ candidates. Pairs with an invariant mass,
under the charged-pion hypothesis, $m_{\pi^{+}\pi^{-}}$ between $635\,\MeV$
and $915\,\MeV$ are selected as $\rho\to\PiPi$ candidates. The
candidates where $m_{K^{+}K^{-}}$ is consistent with the $\phi$ meson
mass are rejected from the $\rho\gamma$ analysis. This requirement
rejects a negligible fraction of the signal in the $\rho\gamma$
analysis.
Selected $M$ candidates are required to satisfy an isolation
requirement: the sum of the $\pt$ of the reconstructed ID tracks from
the primary vertex within $\Delta R= 0.2$ of the leading charged
hadron candidate (excluding the charged-hadron candidates defining
the pair) is required to be less than 10\% of the $\pt$ of the $M$
candidate.

The $M$ candidates are combined with the photon candidates, to form
$M\gamma$ candidates. When multiple combinations are possible, a
situation that arises only in a few percent of the events, the
combination of the highest-\pt photon and the $M$ candidate with an
invariant mass closest to the respective meson mass is selected. The
event is retained for further analysis if the requirement
$\Delta\phi(M,\gamma)>\pi/2$ is satisfied.
The transverse momentum of the $M$ candidates is required to be
greater than a threshold that varies as a function of the invariant
mass of the three-body system, $m_{M\gamma}$.
Thresholds of $40\,\GeV$ and $47.2\,\GeV$ are imposed on $\pt^{M}$ for the regions $m_{M\gamma}<91\,\GeV$ and
$m_{M\gamma}\geq140\,\GeV$, respectively. The threshold is varied from
$40\,\GeV$ to $47.2\,\GeV$ as a linear function of $m_{M\gamma}$ in the
region $91\leq m_{M\gamma}<140\,\GeV$. This approach ensures good
sensitivity for both the Higgs and $Z$ boson searches, while keeping a
single kinematic selection.

For the $\phi(\to\KK)\,\gamma$ final state, the total signal efficiencies
(kinematic acceptance, trigger and reconstruction efficiencies) are
17\% and 10\% for the Higgs and $Z$~boson decays, respectively. The
corresponding efficiencies for the $\rho\gamma$ final state are
8\% and 2.4\%.
The difference in efficiency between the Higgs and $Z$ boson decays
arises primarily from the softer $\pt$ distributions of the photon
and charged-hadron candidates associated with the $Z\to M\,\gamma$
production, as can be seen for the $\phi\gamma$ case by comparing
Figures~\ref{fig:pt_ph_kaons_H} and \ref{fig:pt_ph_kaons_Z}.
The overall lower efficiency in the $\rho\gamma$ final state is
a result of the lower efficiency of the $m_{M}$ requirement due to the
large $\rho$-meson natural width and the different kinematics of the $\rho$
decay products%
, as presented in Figures~\ref{fig:pt_ph_pions_H_rho} and~\ref{fig:pt_ph_pions_Z_rho}.
Meson helicity effects have a relatively small impact for the
$\phi\to\KK$ decays, where the kaons carry very little momentum in the
$\phi$ rest frame. Specifically, the expected Higgs ($Z$) boson signal
yield in the signal region is 2.4\% larger (9\% larger) than in the hypothetical scenario where the meson is unpolarised. For the
$\rho\to\PiPi$ decays the yields are increased by 12\% (increased
by 7\%).

\begin{figure}[!h]
\centering
\subfigure[\label{fig:pt_ph_kaons_H}]{\includegraphics[width=0.40\textwidth]{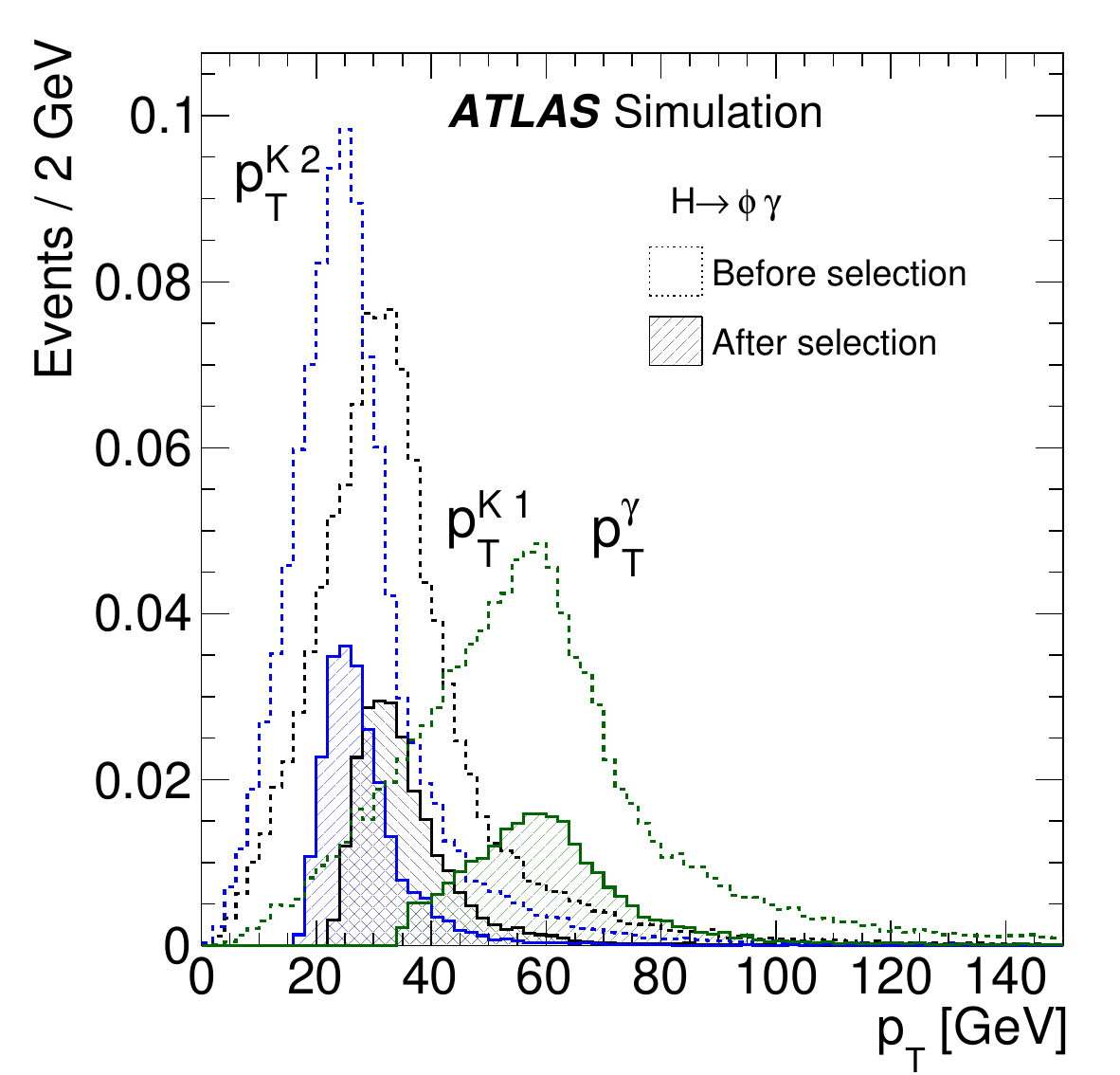}}
\subfigure[\label{fig:pt_ph_kaons_Z}]{\includegraphics[width=0.40\textwidth]{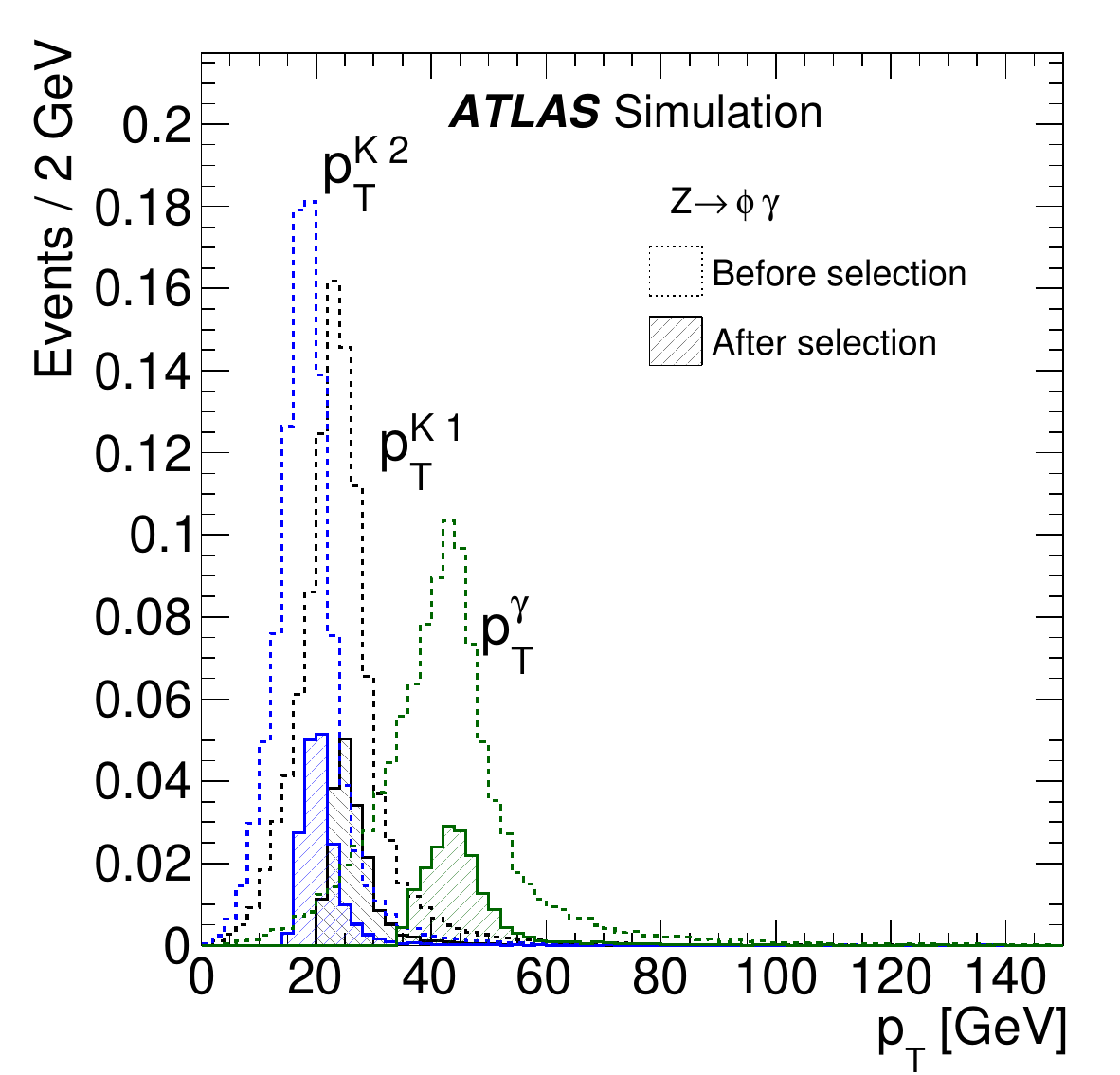}}
\vspace{-0.4cm}

\subfigure[\label{fig:pt_ph_pions_H_rho}]{\includegraphics[width=0.40\textwidth]{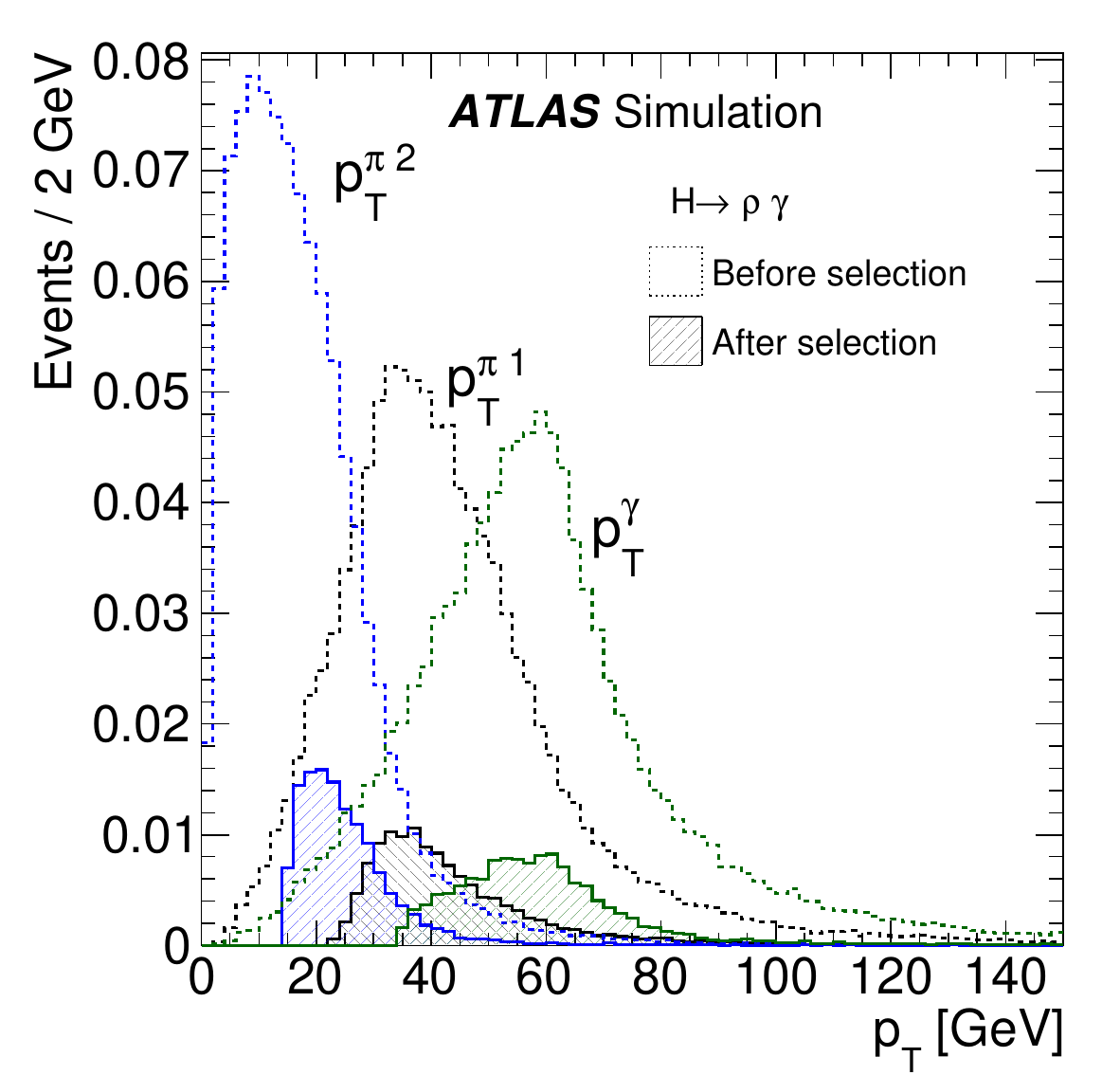}}
\subfigure[\label{fig:pt_ph_pions_Z_rho}]{\includegraphics[width=0.40\textwidth]{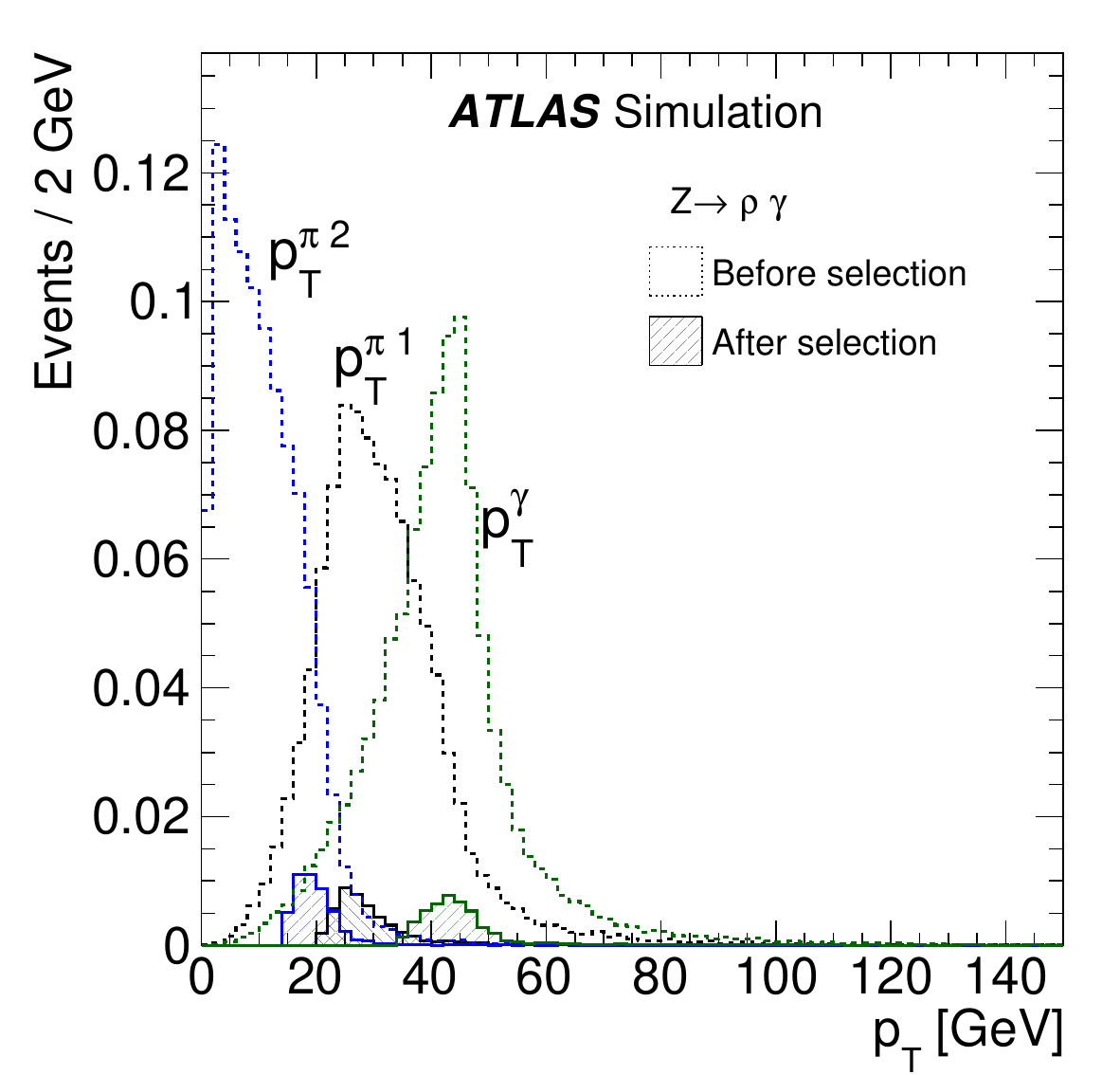}}
\vspace{-0.4cm}
\caption{Generator-level transverse momentum ($\pt$) distributions of the photon
and of the charged-hadrons, ordered in $\pt$, for \subref{fig:pt_ph_kaons_H}
$H\to\phi\gamma$, \subref{fig:pt_ph_kaons_Z} $Z\to\phi\gamma$,
\subref{fig:pt_ph_pions_H_rho} $H\to\rho\gamma$ and
\subref{fig:pt_ph_pions_Z_rho} $Z\to\rho\gamma$ simulated signal events, respectively. The hatched histograms denote the full event selection while the
dashed histograms show the events at generator level that fall within the analysis
geometric acceptance (both charged-hadrons are required to have $|\eta|<2.5$ while the
photon is required to have $|\eta|<2.37$, excluding the region
$1.37<|\eta|<1.52$). The dashed histograms are normalised to unity, and the
relative difference between the two sets of distributions corresponds to the
effects of reconstruction, trigger, and event selection efficiencies. The
leading charged-hadron candidate $h= K, \pi$ is denoted by $p_\textrm{T}^{h1}$
and the sub-leading candidate by $p_\textrm{T}^{h2}$.\label{fig:pt_ph_kaons}}
\end{figure}

The average $m_{M\gamma}$ resolution is 1.8\% for both the Higgs and $Z$~boson
decays. The Higgs boson signal $m_{M\gamma}$ distribution is modelled with a sum of two Gaussian
probability density functions (pdf) with a common mean value, while
the $Z$ boson signal $m_{M\gamma}$ distribution is modelled with a double Voigtian pdf (a convolution of
 relativistic Breit--Wigner and Gaussian pdfs) corrected with a mass-dependent
efficiency factor.

The $m_{\KK}$ distribution for the selected $\phi\gamma$ candidates,
with no $m_{\KK}$ requirement applied, is shown in
Figure~\ref{fig:MassPhiFit} exhibiting a visible peak at the $\phi$
meson mass. The $\phi$ peak is fitted with a Voigtian pdf, while the
background is modelled with a function typically used to
describe kinematic thresholds~\cite{Lees:2013uxa}. The experimental
resolution in $m_{K^{+}K^{-}}$ is approximately $4\,\MeV$, comparable to the
$4.3\,\MeV$~\cite{PDG} width of the $\phi$ meson.
In Figure~\ref{fig:MassRhoFit}, the corresponding distribution for the
selected $\rho\gamma$ candidates is shown, where the $\rho$ meson
can also be observed. The $\rho$ peak is fitted with a single Breit--Wigner pdf,
modified by a mass-dependent width to match the distribution obtained from \PYTHIA~\cite{Sjostrand:2007gs}. The
background is fitted with the sum of a combinatoric background, estimated from events containing a
same-sign di-track pair, and other backgrounds determined in the fit
using a linear combination of Chebychev polynomials up to the second order.
Figure~\ref{fig:MassFit} only qualitatively illustrates
the meson selection in the studied final state, and is not used
any further in this analysis.

\begin{figure}[h]
\centering
\subfigure[\label{fig:MassPhiFit}]{\includegraphics[width=0.45\linewidth]{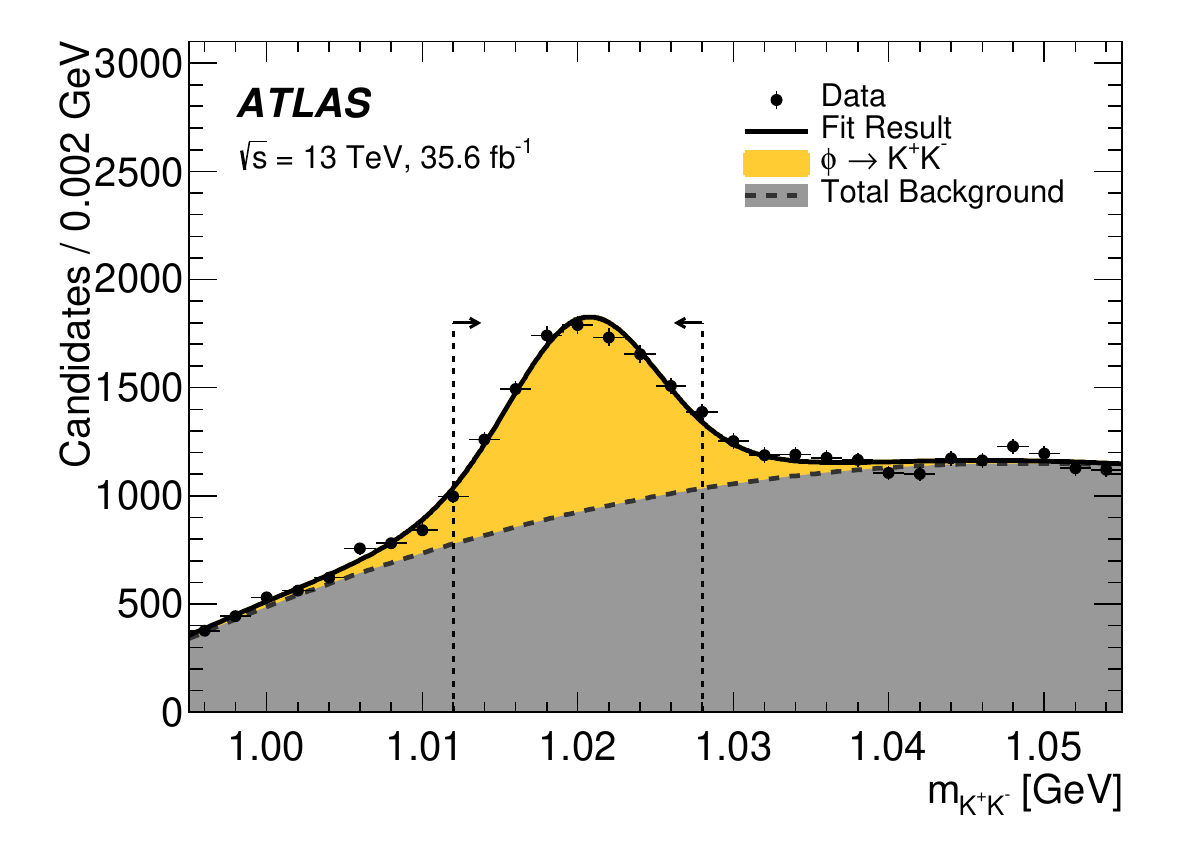}}
\subfigure[\label{fig:MassRhoFit}]{\includegraphics[width=0.45\linewidth]{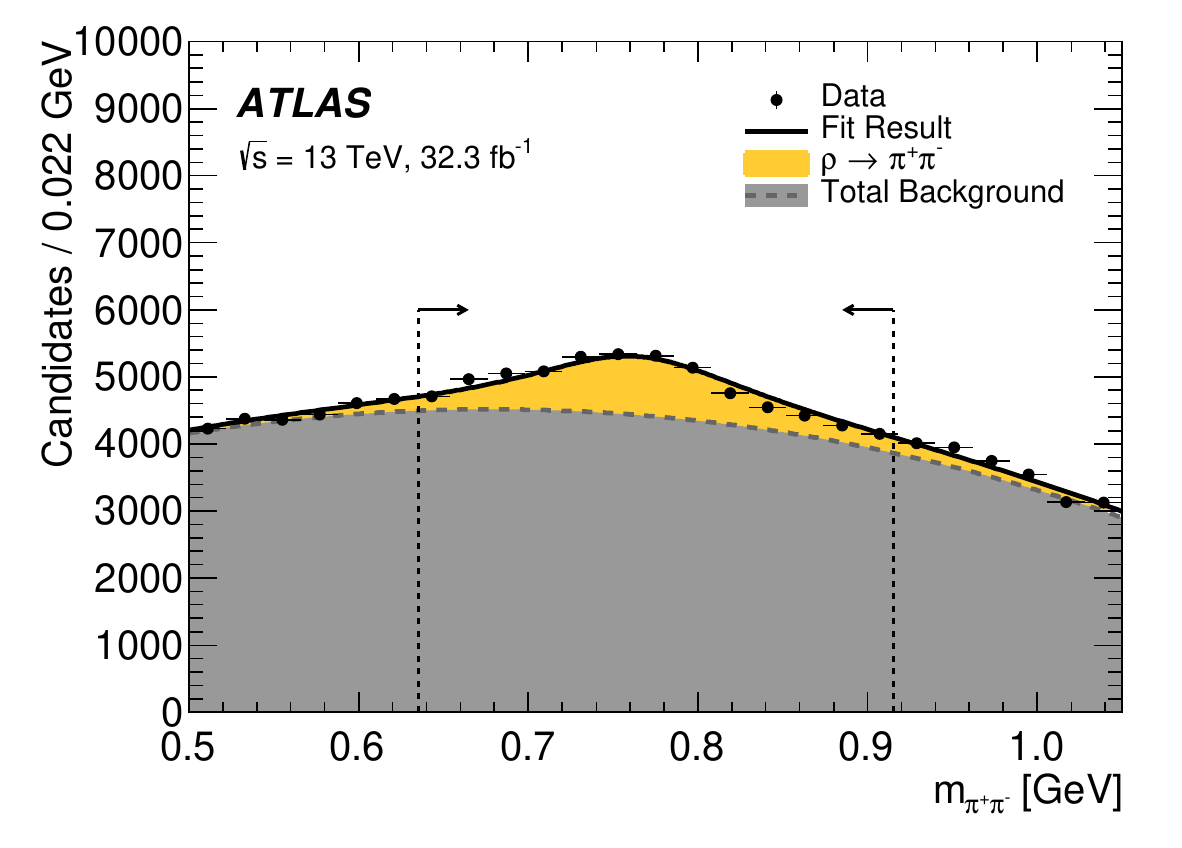}}
\vspace{0.2cm}
\caption{The \subref{fig:MassPhiFit} $m_{\KK}$ and \subref{fig:MassRhoFit} $m_{\PiPi}$
 distributions for $\phi\gamma$ and $\rho\gamma$
candidates, respectively. The candidates fulfil the complete event selection
(see text), apart from requirements on $m_{\KK}$ or $m_{\PiPi}$. These
requirements are marked on the figures with dashed lines topped with arrows
indicating the included area. The signal and background models are discussed in
the text.
\label{fig:MassFit}}
\end{figure}

\section{Background}
\label{sec:background}
For both the $\phi\gamma$ and $\rho\gamma$ final states, the main
sources of background in the searches are events involving inclusive
photon + jet or multijet processes where an $M$ candidate is
reconstructed from ID tracks originating from a jet.

From the selection criteria discussed earlier, the shape of this
background exhibits a turn-on structure in the $m_{M\gamma}$
distribution around $100\,\GeV$, in the region of the $Z$ boson signal,
and a smoothly falling background in the region of the Higgs
boson signal. Given the complex shape of this background,
these processes are modelled in an inclusive fashion with a
non-parametric data-driven approach using templates to describe the
relevant distributions. The background normalisation and
shape are simultaneously
extracted from a fit to the data. A similar procedure was used in the
earlier search for Higgs and $Z$ boson decays into
$\phi\gamma$~\cite{HIGG-2016-05} and the search for Higgs and $Z$
boson decays into $\Jpsi\,\gamma$ and $\Upsilon(nS)\,\gamma$ described
in Ref.~\cite{HIGG-2014-03}.

\subsection{Background modelling}
\label{sec:backgroundmodel}

The background modelling procedure for each final state exploits a
sample of approximately 54\,000 $\KK\gamma$ and 220\,000 $\PiPi\gamma$
candidate events in data. These events pass all the
kinematic selection requirements described previously, except that the
photon and $M$ candidates are not required to satisfy the nominal
isolation requirements, and a looser $\pt^{M}>35\,\GeV$ requirement is
imposed. This selection defines the background-dominated ``generation
region'' (GR).
From these events, pdfs are constructed to describe the distributions
of the relevant kinematic and isolation variables and their most
important correlations. In this way, in the absence of appropriate simulations,
pseudocandidate events are generated, from which the
background shape in the discriminating variable is derived.

This ensemble of pseudocandidate events is produced by randomly
sampling the distributions of the relevant kinematic and isolation
variables, which are estimated from the data in the GR. Each
pseudocandidate event is described by $M$ and $\gamma$
four-momentum vectors and the associated $M$ and photon isolation
variables. The $M$ four-momentum vector is constructed from sampled $\eta_{M}$,
$\phi_{M}$, $m_{M}$ and $\pt^{M}$ values. For the $\gamma$
four-momentum vector, the $\eta_{\gamma}$ and $\phi_{\gamma}$ are determined from the
sampled $\Delta\phi(M,\gamma)$ and $\Delta\eta(M,\gamma)$ values whereas 
$\ptgamma$ is sampled directly.

The most important correlations among these kinematic
and isolation variables in background events are retained in the
generation of the pseudocandidates through the following 
sampling scheme, where the steps are performed sequentially:

\begin{enumerate}[i)]
  \item Values for $\eta_{M}$, $\phi_{M}$, $m_{M}$ and $\pt^{M}$ are drawn randomly and independently according to the corresponding pdfs.
  \item The distribution of $\ptgamma$ values is parameterised in bins of $\pt^{M}$, and values are drawn from the corresponding bins given the previously generated value of $\pt^{M}$.
The $M$ isolation variable is parameterised in bins of
$\pt^{M}(\ptgamma)$\ for the $\phi\gamma$ ($\rho\gamma$) model and sampled accordingly. The difference between the two approaches for the $\phi\gamma$\ and
$\rho\gamma$ accounts for the difference in the observed correlations
arising in the different datasets.

\item The distributions of the values for $\Delta\eta(M,\gamma)$, photon calorimeter isolation, normalised to $\ptgamma$, and their correlations are parameterised in a 
two-dimensional distribution. For the $\phi\gamma$ analysis, 
several distributions are produced corresponding to
the $\pt^{M}$ bins used earlier to describe the $\ptgamma$ and $M$ isolation variables, whereas for the $\rho\gamma$ final state the 
two-dimensional distribution is produced inclusively for all $\pt^{M}$ values.

\item The photon track isolation, normalised to $\ptgamma$, and the $\Delta\phi(M,\gamma)$ variables are sampled from pdfs generated in bins of relative photon calorimeter
  isolation and $\Delta\eta(M,\gamma)$, respectively, using the values drawn in step iii).
\end{enumerate}

The nominal selection requirements are imposed on the ensemble, and the
surviving pseudocandidates are used to construct templates for the
$m_{M\gamma}$ distribution, which are then smoothed using Gaussian
kernel density estimation~\cite{Cranmer:2000du}.
It was verified through signal
injection tests that the shape of the background model is not affected by potential signal contamination.

\subsection{Background validation}
\label{sec:backgroundvalid}

To validate the background model, the $m_{M\gamma}$
distributions in several validation regions, defined by kinematic and
isolation requirements looser than the nominal signal requirements,
are used to compare the prediction of the background model with the
data. Three validation regions are defined, each based on the GR
selection and adding one of the following: the $\pt^{M}$
requirement (VR1), the photon isolation requirements (VR2), or the
meson isolation requirement (VR3).
The $m_{M\gamma}$ distributions in these validation regions are shown
in Figure~\ref{fig:BkgdVal}.
The background model is found to describe the data in all 
regions within uncertainties (see Section~\ref{sec:systematic}).

Potential background contributions from $Z\to\ell\ell\gamma$
decays and inclusive Higgs decays were studied and found to
be negligible for the selection requirements and
dataset used in this analysis.

\begin{figure}[!htbp]
  \begin{center}
    \includegraphics[width=0.32\textwidth]{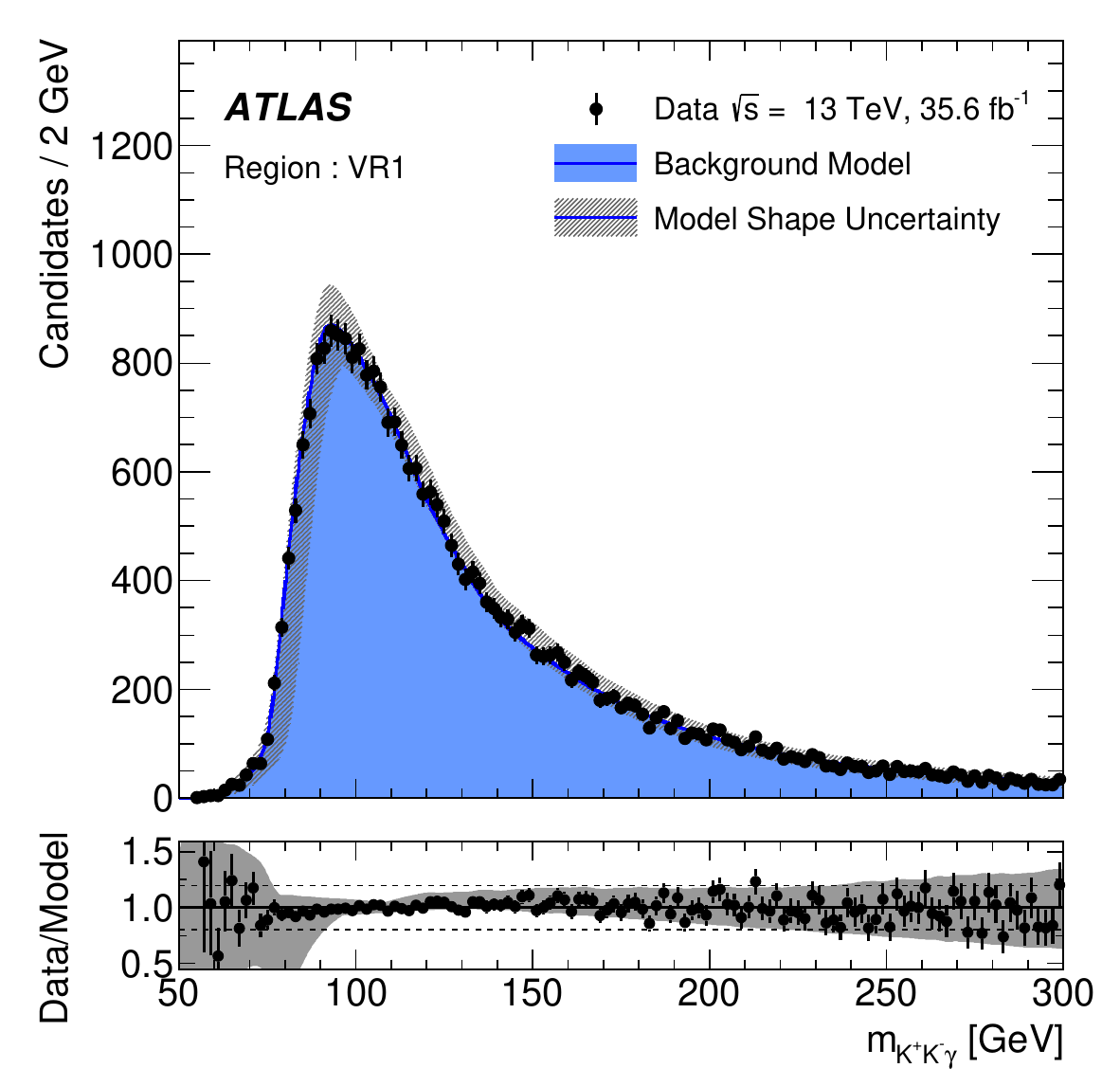}
    \includegraphics[width=0.32\textwidth]{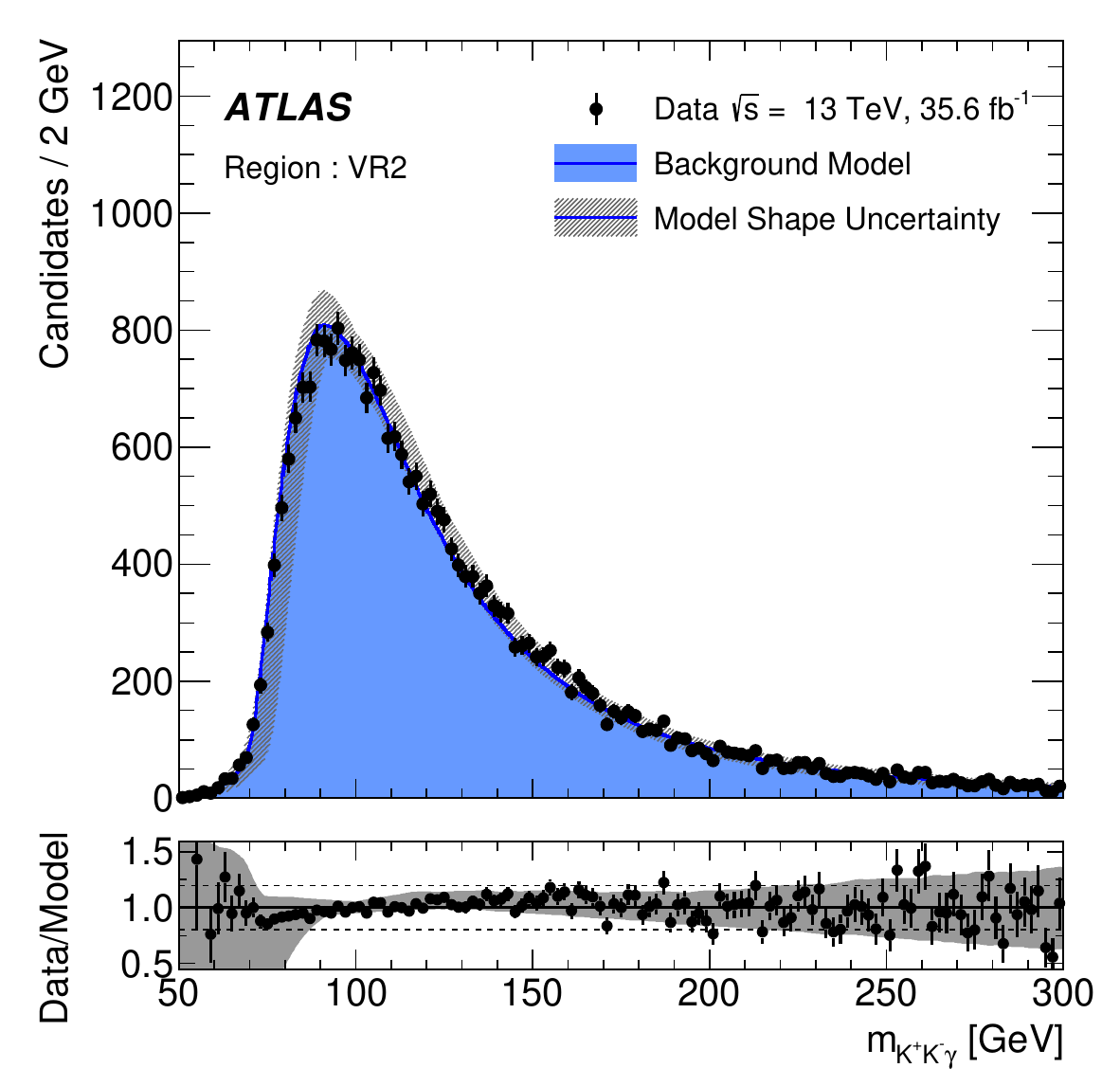}
    \includegraphics[width=0.32\textwidth]{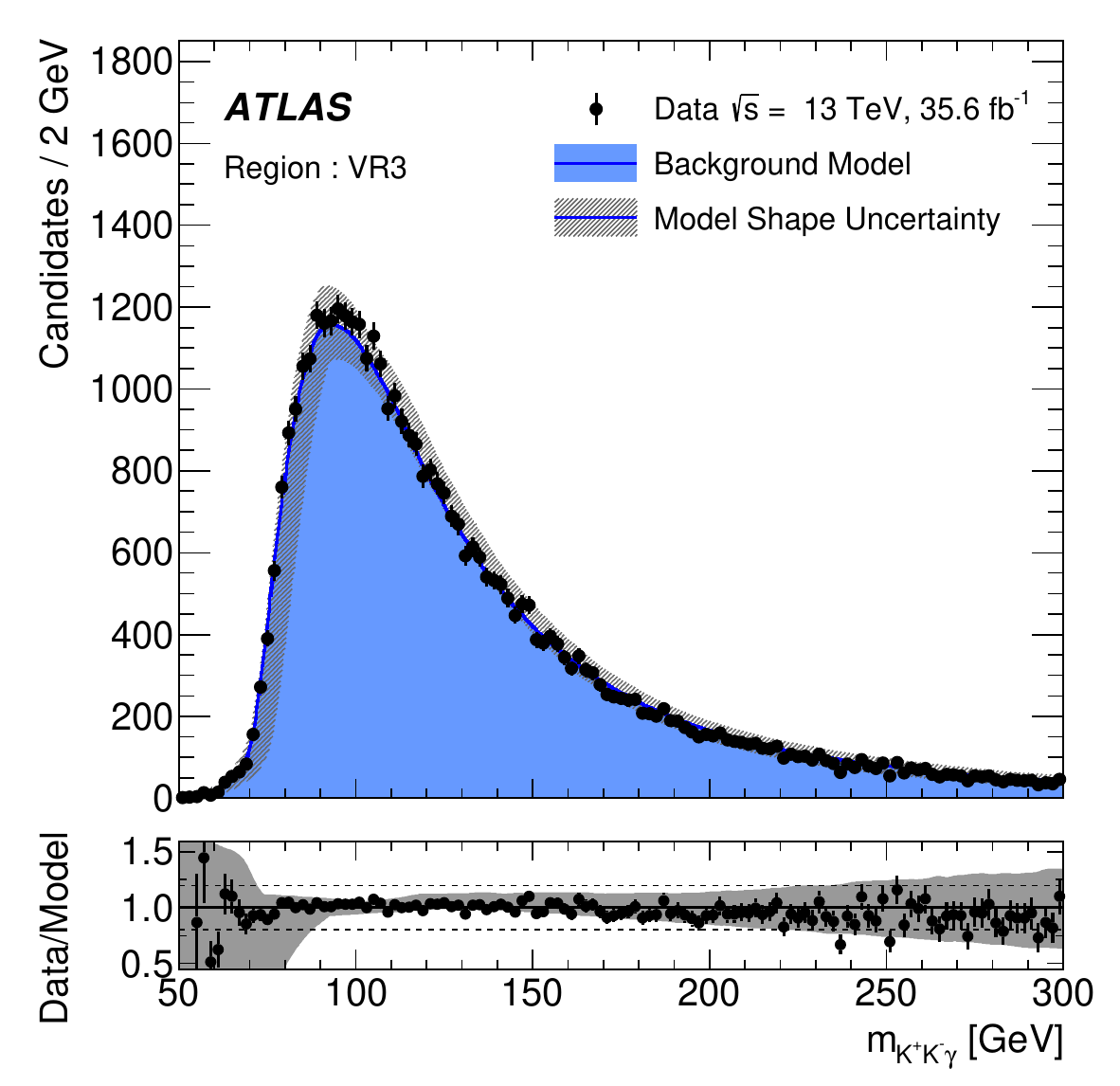}

    \includegraphics[width=0.32\textwidth]{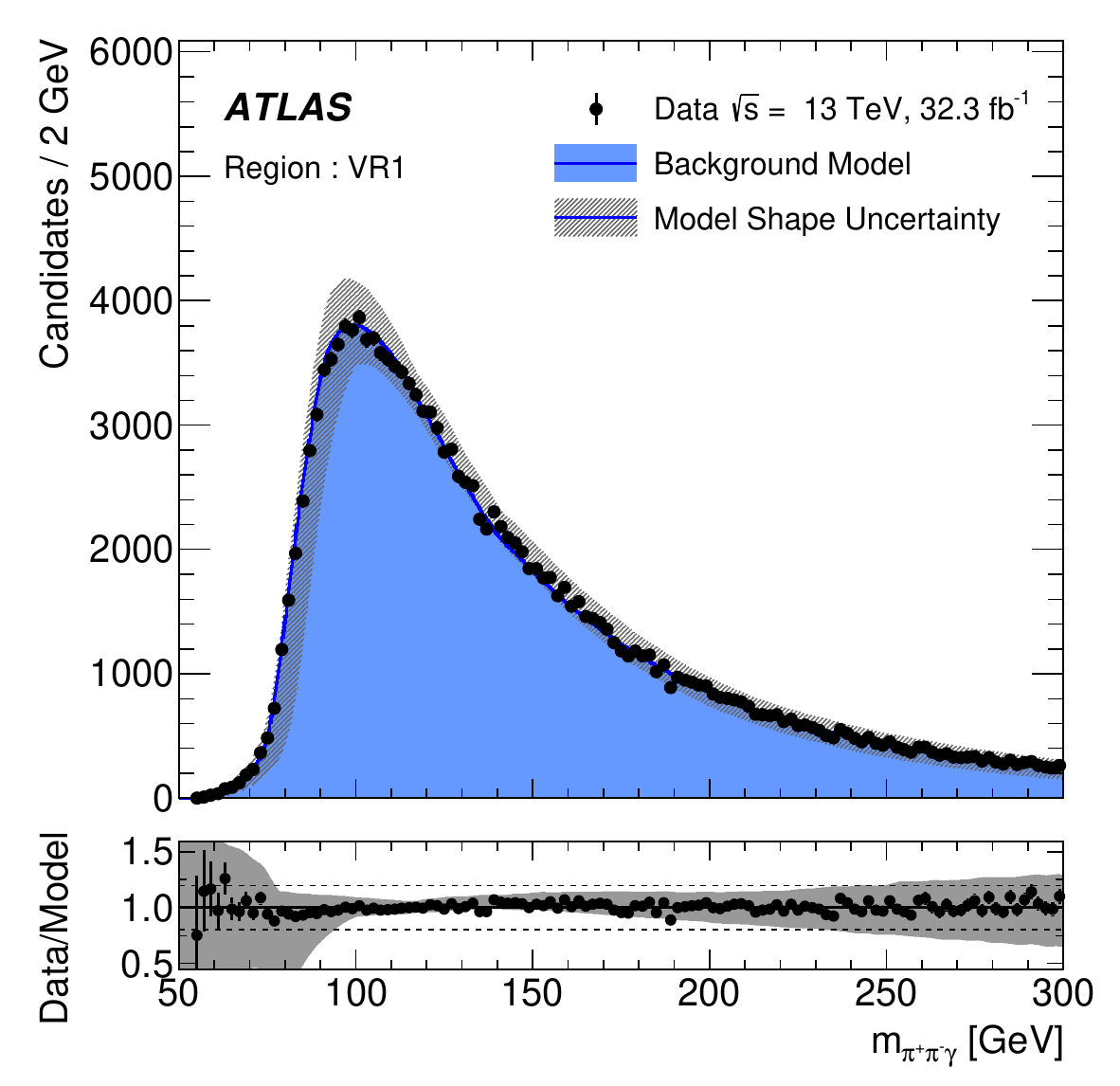}
    \includegraphics[width=0.32\textwidth]{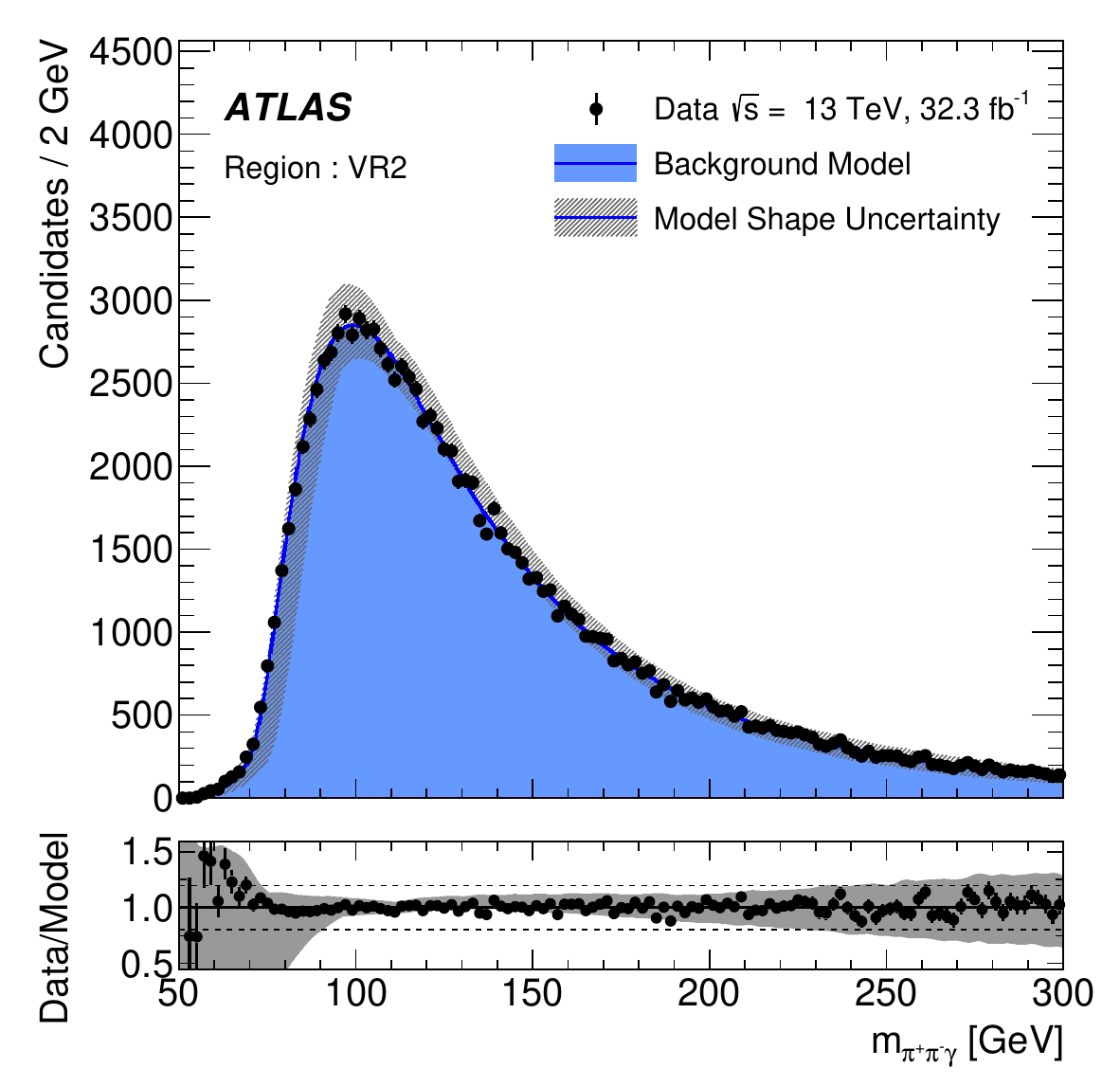}
    \includegraphics[width=0.32\textwidth]{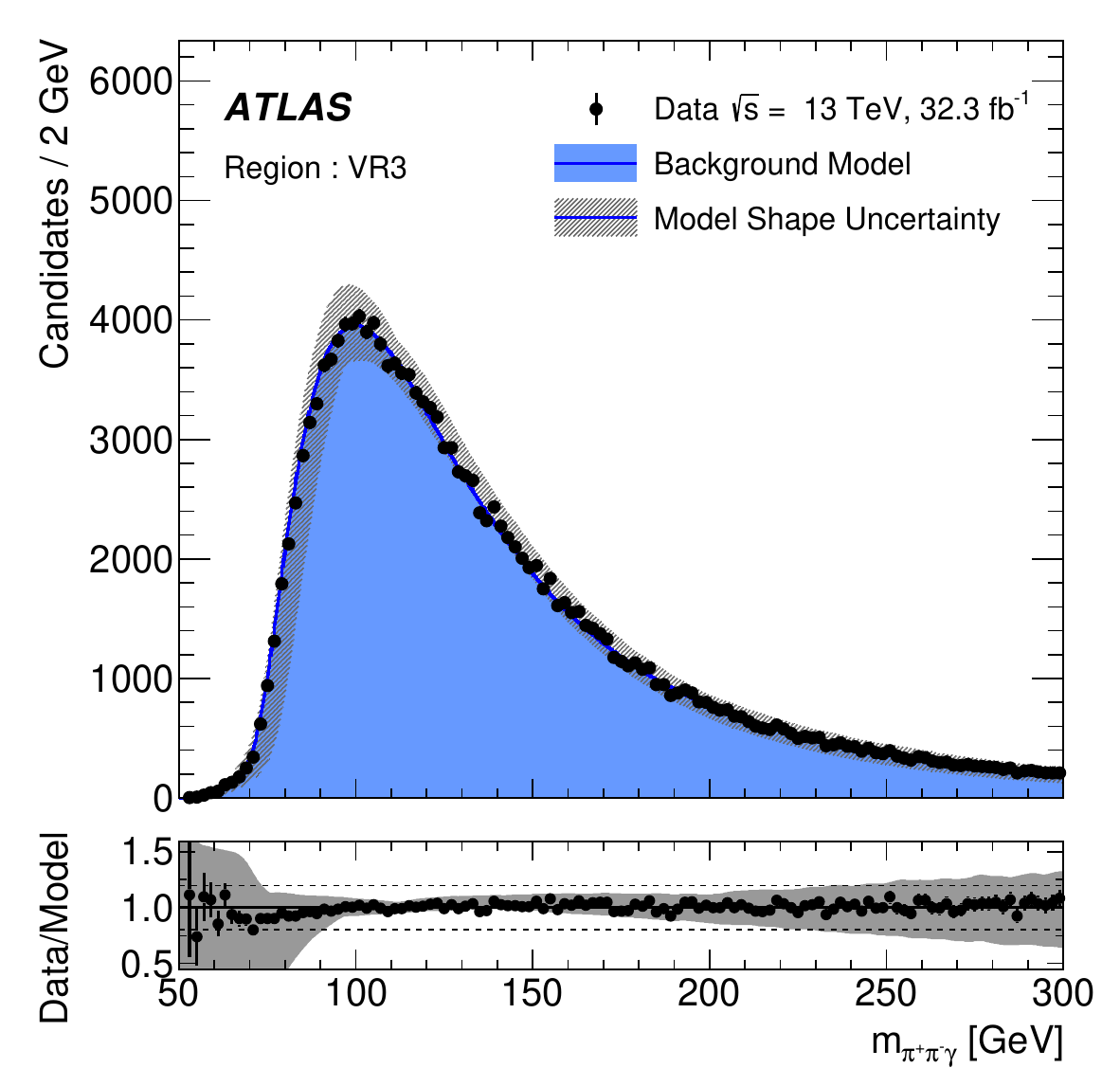}
  \end{center}
\vspace{-0.6cm}
\caption{The distribution of $m_{\KK\gamma}$ top ($m_{\PiPi\gamma}$ bottom) in data compared to the prediction of the background model for the VR1, VR2 and VR3 validation regions. The background model is normalised to the observed number of events within the region shown. The uncertainty band corresponds to the uncertainty envelope
derived from variations in the background modelling procedure. The ratio of the data to the background model is shown below the distributions.\label{fig:BkgdVal}
}
\end{figure}

A further validation of the background modelling is performed using
events within a sideband of the $M$ mass distribution. For the
$\phi\gamma$ analysis the sideband region is defined by $1.035\,\gev <
\mKK < 1.051\,\gev$. For the $\rho\gamma$ analysis the sideband region
is defined by $950\,\MeV < m_{\PiPi} < 1050\,\MeV$. All other selection
requirements and modelling procedures are identical to those  used in the
signal region. Figures \ref{fig:BkgdSidebandsPhi} and
\ref{fig:BkgdSidebandsRho} show the $m_{M\gamma}$ distributions
for the sideband region. The background model is found to describe the
data within the systematic uncertainties described in
Section~\ref{sec:systematic}.

\begin{figure}[htb]
  \begin{center}
    \subfigure[\label{fig:BkgdSidebandsPhi}]{\includegraphics[width=0.45\textwidth]{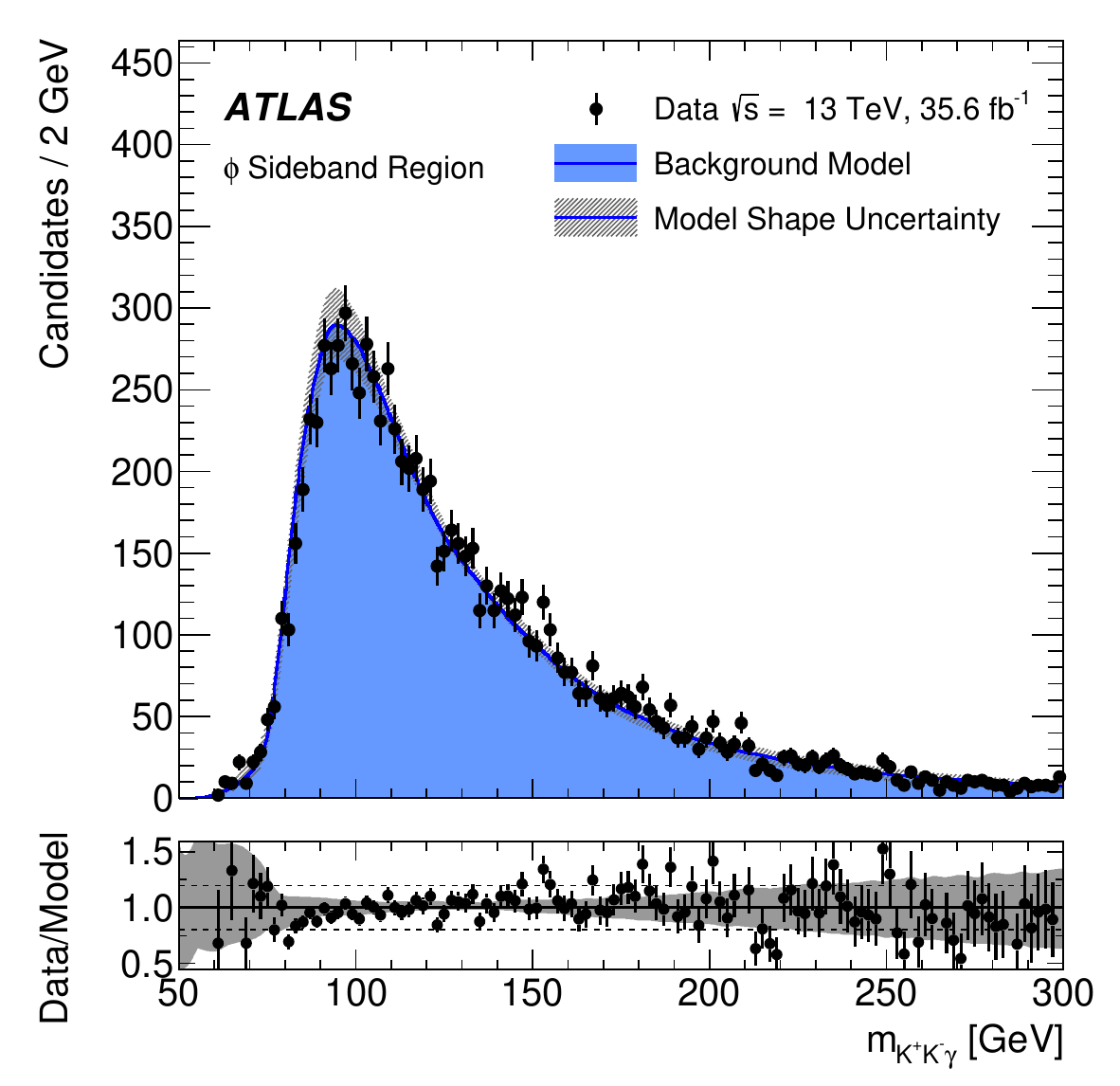}}
    \subfigure[\label{fig:BkgdSidebandsRho}]{\includegraphics[width=0.45\textwidth]{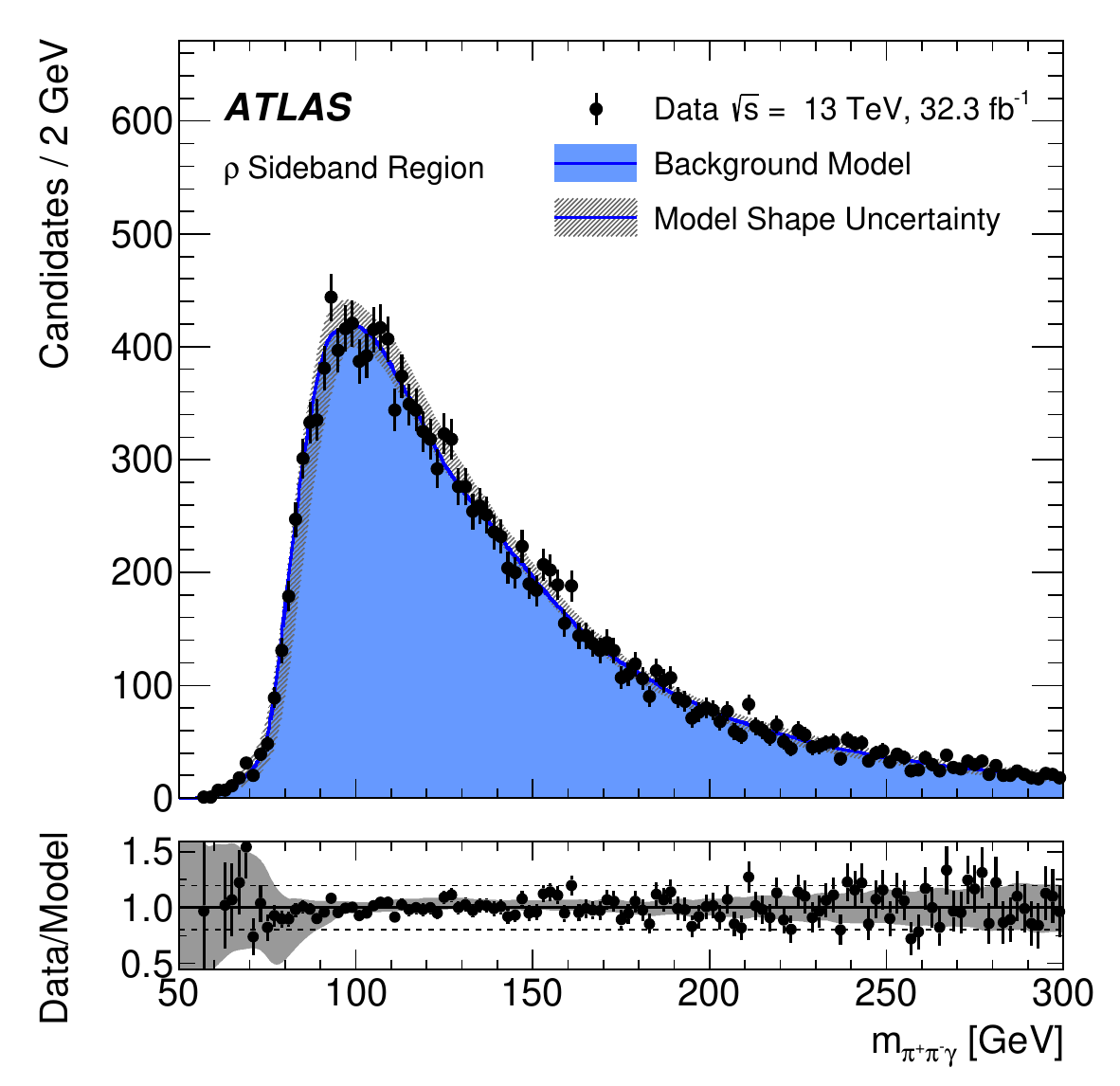}}
  \end{center}
\vspace{-0.6cm}
  \caption{The distribution of $m_{M\gamma}$ for the 
  \subref{fig:BkgdSidebandsPhi} $\phi\gamma$\ and \subref{fig:BkgdSidebandsRho}
  $\rho\gamma$\ selections in the sideband control region. The background
  model is normalised to the observed number of events within the region shown.
  The uncertainty band corresponds to the uncertainty envelope derived from
  variations in the background modelling procedure. The ratio of the data to the background model is shown below the distributions.\label{fig:BkgdSidebands}}
\end{figure}

\FloatBarrier
\section{Systematic uncertainties}
\label{sec:systematic}

Trigger and identification efficiencies for photons are determined
from samples enriched with $Z\to\ee$ events in
data~\cite{PERF-2013-04,TRIG-2016-01}. The systematic
uncertainty in the expected signal yield associated with the trigger
efficiency is estimated to be 2.0\%. The photon identification
and isolation uncertainties, for both the converted and unconverted photons,
are estimated to be 2.4\% and 2.6\% for the Higgs and $Z$~boson
signals, respectively. An uncertainty of 6.0\% per $M$ candidate is
assigned to the track reconstruction efficiency and accounts for
effects associated with the modelling of ID material and track reconstruction
algorithms if a nearby charged
particle is present. This uncertainty is derived conservatively by
assuming a 3\% uncertainty in the reconstruction efficiency of each
track~\cite{PERF-2015-08}, and further assuming the uncertainty to
be fully correlated between the two tracks of the $M$ candidate.

The systematic uncertainties in the Higgs production cross section are obtained
from Ref.~\cite{deFlorian:2016spz} as described in
Section~\ref{sec:datamc}. The $Z$ boson production cross-section uncertainty is
taken from the measurement in Ref.~\cite{STDM-2015-03}.

The photon energy scale uncertainty, determined from $Z\to\ee$ events
and validated using $Z\to\ell\ell \gamma$
events~\cite{PERF-2013-05}, is applied to the
simulated signal samples as a function of $\etagamma$ and $\ptgamma$.
The impact of the photon energy scale uncertainty on the Higgs and $Z$
boson mass distributions does not exceed 0.2\%. The uncertainty associated with the
photon energy resolution is found to have a negligible
impact. Similarly, the systematic uncertainty associated with the ID
track momentum measurement is found to be negligible. The systematic
uncertainties in the expected signal yields
are summarised in Table~\ref{tab:systematics}.

The shape of the background model is allowed to vary around the
nominal shape, and the parameters controlling these systematic variations
are treated as nuisance parameters in the maximum-likelihood fit used to
extract the signal and background yields.
Three such shape variations are implemented through
varying $\ptgamma$, linear
distortions of the shape of the $\Delta\phi({M},\gamma)$, and a
global tilt of the three-body mass.
The first two variations alter the kinematics
of the pseudocandidates that are propagated to the
three-body mass.

\begin{table}[h]
\caption{Summary of the relative systematic uncertainties in the expected signal yields. The magnitude of the effects are the same for both the $\phi\gamma$ and $\rho\gamma$ selections.\label{tab:systematics}}
\centering
\begin{tabular}{lcc}
  \hline
  Source of systematic uncertainty & Yield uncertainty \\\hline
Total $H$ cross section & $6.3\%$ \\
Total $Z$ cross section & $2.9\%$ \\
Integrated luminosity   & $3.4\%$ \\
Photon ID efficiency    & $2.5\%$\\ 
Trigger efficiency      & $2.0\%$ \\ 
Tracking efficiency     & $6.0\%$  \\ \hline
\end{tabular}
\end{table}

\section{Results}
\label{sec:results}
The data are compared to background and signal predictions using an
unbinned maximum-likelihood fit to the $m_{M\gamma}$ distribution.
The parameters of interest are the Higgs and $Z$ boson
signal normalisations. Systematic uncertainties are modelled using
additional nuisance parameters in the fit; in particular the
background normalisation is a free parameter in the model. 
The fit uses the selected events with $m_{{M}\gamma}<300\,\gev$.
The expected and observed numbers of background events within the
$m_{M\gamma}$ ranges relevant to the Higgs and $Z$ boson signals are
shown in Table~\ref{tab:events}. The observed yields are consistent
with the number of events expected from the background-only prediction
within the systematic and statistical uncertainties.
The results of the background-only fits for the $\phi\gamma$ and
$\rho\gamma$ analyses are shown in Figures~\ref{fig:FitPhi} and
\ref{fig:FitRho}, respectively.

\begin{table}[h]
\caption{The number of observed events and the mean expected
  background, estimated from the maximum-likelihood fit and shown with
  the associated total uncertainty, for the $m_{M\gamma}$ ranges of
  interest. The expected Higgs and $Z$~boson signal yields, along with
  the total systematic uncertainty, for $\phi\gamma$ and $\rho\gamma$,
  estimated using simulations, are also shown in parentheses.
  \label{tab:events}}
\centering
\begin{tabular}{c|r|rr|rr|c|c}
\hline
&\multicolumn{5}{c}{Observed yields (Mean expected background)}& \multicolumn{2}{|c}{Expected signal yields} \\ \cline{2-8}
&\multicolumn{5}{c|}{Mass range [GeV]}  & $H$ & $Z$\\ \cline{2-6}
             & All & \multicolumn{2}{c|}{$81$--$101$} & \multicolumn{2}{c|}{$120$--$130$}  & $[{\cal B} = 10^{-4}]$ & $[{\cal B} = 10^{-6}]$ \\ \hline     
             $\phi\gamma$ & 12051  & 3364  & (3500 $\pm$ 30)  & 1076 & (1038 $\pm$ \hphantom{0}9)  & 15.1 $\pm$ 1.5 & 98 $\pm$ 8  \\
             $\rho\gamma$ & 58702  & 12583 & (12660 $\pm$ 60) & 5473 & (5450 $\pm$ 30)             & 14.3 $\pm$ 1.4 & 47 $\pm$ 4  \\
\hline
\end{tabular}
\end{table}

\begin{figure}[h]
\centering
\subfigure[\label{fig:FitPhi}]{\includegraphics[width=0.45\linewidth]{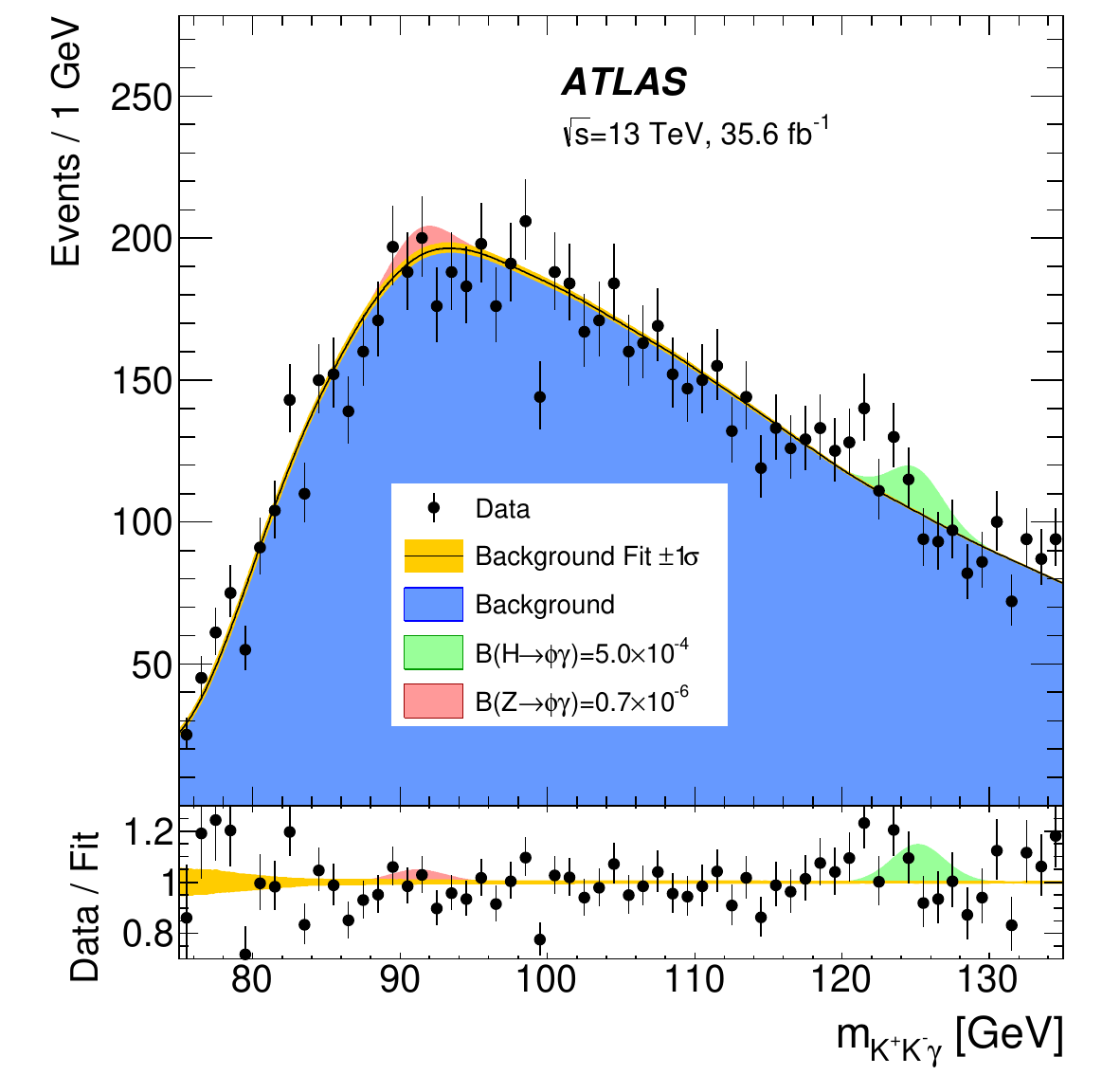}}
\subfigure[\label{fig:FitRho}]{\includegraphics[width=0.45\linewidth]{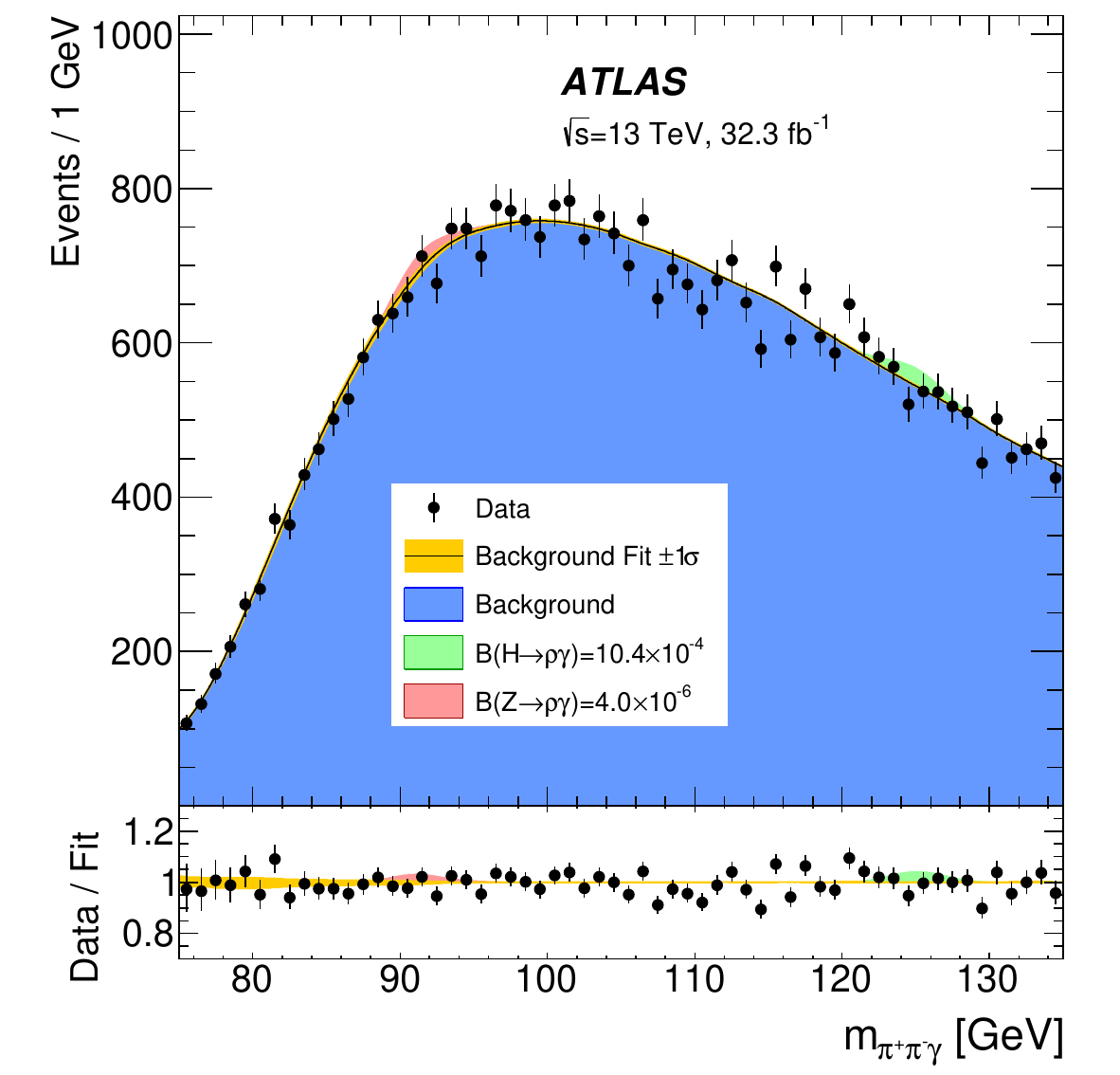}}
\vspace{-0.4cm}
\caption{The \subref{fig:FitPhi} $m_{\KK\gamma}$ and \subref{fig:FitRho} $m_{\PiPi\gamma}$ distributions of the
selected $\phi\gamma$ and $\rho\gamma$ candidates, respectively, along with
the results of the maximum-likelihood fits with a background-only model. The
Higgs and $Z$~boson contributions for the branching fraction values
corresponding to the observed 95\% CL upper limits are also shown. Below the figures the ratio of the data to the background-only fit is shown.
\label{fig:Fit}}
\end{figure}

Upper limits are set on the
branching fractions for the Higgs and $Z$~boson decays into $M\,\gamma$
using the CL$_{\text{s}}$ modified frequentist formalism~\cite{cls} with the
profile-likelihood-ratio test statistic~\cite{Cowan:2010js}. For the
upper limits on the branching fractions, the SM production
cross section is assumed for the Higgs boson~\cite{deFlorian:2016spz},
while the ATLAS measurement of the inclusive $Z$~boson cross section
is used for the $Z$~boson signal~\cite{STDM-2015-03}, as discussed in
Section~\ref{sec:datamc}. The results are summarised in
Table~\ref{fig:Limit}.
The observed 95\% CL upper limits on the branching
fractions for $H\to\phi\gamma$ and $Z\to\phi\,\gamma$ decays are
$208$ and $87$ times the expected SM branching fractions,
respectively. The corresponding values for the $\rho\gamma$ decays
are $52$ and $597$ times the expected SM branching fractions,
respectively.
Upper limits at 95\% CL on the production cross section
times branching fraction are also estimated for the Higgs boson
decays, yielding \SI{26.1}{\femto\barn} for the $H\to\phi\gamma$ decay, and
\SI{54.8}{\femto\barn} for the $H\to\rho\gamma$ decay.

The systematic uncertainties described in Section~\ref{sec:systematic}
result in a 14\% deterioration of the post-fit expected 95\% CL upper
limit on the branching fraction in the $H\to\phi\gamma$ and $Z\to\phi\gamma$ analyses, compared to the result including only statistical uncertainties.
For the $\rho\gamma$ analysis the systematic uncertainties result in a 2.3\%
increase in the post-fit expected upper limit for the Higgs boson decay, while 
for the $Z$ boson decay the upper limit deteriorates by 29\%. 

\begin{table}[h!]
\caption{Expected and observed branching fraction upper limits at 95\%
  CL for the $\phi\gamma$ and $\rho\gamma$ analyses. The $\pm
  1\sigma$ intervals of the expected limits are also
  given.\label{fig:Limit}} \centering
\begin{tabular}{c|c|c}
\hline
Branching Fraction Limit (95\% CL) & Expected & Observed \\ \hline
${\cal B}\left(H\to\phi\gamma\right)[\tabscript{10}{-4}{}]$ & \hphantom{0}$4.2_{-1.2}^{+1.8}$ & \hphantom{0}5.0 \\
${\cal B}\left(Z\to\phi\gamma\right)[\tabscript{10}{-6}{}]$ & \hphantom{0}$1.1_{-0.3}^{+0.5}$ & \hphantom{0}0.7 \\
${\cal B}\left(H\to\rho\gamma\right)[\tabscript{10}{-4}{}]$ & $10.0_{-2.8}^{+4.9}$ & 10.4 \\
${\cal B}\left(Z\to\rho\gamma\right)[\tabscript{10}{-6}{}]$ & \hphantom{0}$5.1_{-1.4}^{+2.1}$ & \hphantom{0}4.0 \\
\hline
\end{tabular}
\end{table}

\FloatBarrier
\section{Summary}
\label{sec:summary}
A search for the decays of Higgs and $Z$ bosons into $\phi\gamma$ and
$\rho\gamma$ has been performed with $\rts=13\,\TeV$ $pp$ collision data
samples collected with the ATLAS detector at the LHC corresponding to
integrated luminosities of up to $35.6\,\ifb$. The $\phi$ and $\rho$ mesons are
reconstructed via their dominant decays into the $\KK$ and $\PiPi$ final
states, respectively.  The background model is derived using a fully data
driven approach and validated in a number of control regions including
sidebands in the $\KK$ and $\PiPi$ mass distributions.

No significant excess of events above the background
expectations is observed, as expected from the SM. The obtained 95\% CL upper limits are ${\cal
  B}\left(H\to\phi\gamma\right)<5.0\times10^{-4}$, ${\cal
  B}\left(Z\to\phi\gamma\right) < 0.7\times10^{-6}$
,${\cal B}\left(H\to\rho\gamma\right)<10.4\times10^{-4}$ and ${\cal
  B}\left(Z\to\rho\gamma\right) < 4.0\times10^{-6}$.

\section*{Acknowledgements}

We thank CERN for the very successful operation of the LHC, as well as the
support staff from our institutions without whom ATLAS could not be
operated efficiently.

We acknowledge the support of ANPCyT, Argentina; YerPhI, Armenia; ARC, Australia; BMWFW and FWF, Austria; ANAS, Azerbaijan; SSTC, Belarus; CNPq and FAPESP, Brazil; NSERC, NRC and CFI, Canada; CERN; CONICYT, Chile; CAS, MOST and NSFC, China; COLCIENCIAS, Colombia; MSMT CR, MPO CR and VSC CR, Czech Republic; DNRF and DNSRC, Denmark; IN2P3-CNRS, CEA-DRF/IRFU, France; SRNSFG, Georgia; BMBF, HGF, and MPG, Germany; GSRT, Greece; RGC, Hong Kong SAR, China; ISF, I-CORE and Benoziyo Center, Israel; INFN, Italy; MEXT and JSPS, Japan; CNRST, Morocco; NWO, Netherlands; RCN, Norway; MNiSW and NCN, Poland; FCT, Portugal; MNE/IFA, Romania; MES of Russia and NRC KI, Russian Federation; JINR; MESTD, Serbia; MSSR, Slovakia; ARRS and MIZ\v{S}, Slovenia; DST/NRF, South Africa; MINECO, Spain; SRC and Wallenberg Foundation, Sweden; SERI, SNSF and Cantons of Bern and Geneva, Switzerland; MOST, Taiwan; TAEK, Turkey; STFC, United Kingdom; DOE and NSF, United States of America. In addition, individual groups and members have received support from BCKDF, the Canada Council, CANARIE, CRC, Compute Canada, FQRNT, and the Ontario Innovation Trust, Canada; EPLANET, ERC, ERDF, FP7, Horizon 2020 and Marie Sk{\l}odowska-Curie Actions, European Union; Investissements d'Avenir Labex and Idex, ANR, R{\'e}gion Auvergne and Fondation Partager le Savoir, France; DFG and AvH Foundation, Germany; Herakleitos, Thales and Aristeia programmes co-financed by EU-ESF and the Greek NSRF; BSF, GIF and Minerva, Israel; BRF, Norway; CERCA Programme Generalitat de Catalunya, Generalitat Valenciana, Spain; the Royal Society and Leverhulme Trust, United Kingdom.

The crucial computing support from all WLCG partners is acknowledged gratefully, in particular from CERN, the ATLAS Tier-1 facilities at TRIUMF (Canada), NDGF (Denmark, Norway, Sweden), CC-IN2P3 (France), KIT/GridKA (Germany), INFN-CNAF (Italy), NL-T1 (Netherlands), PIC (Spain), ASGC (Taiwan), RAL (UK) and BNL (USA), the Tier-2 facilities worldwide and large non-WLCG resource providers. Major contributors of computing resources are listed in Ref.~\cite{ATL-GEN-PUB-2016-002}.

\printbibliography

\clearpage
 
\begin{flushleft}
{\Large The ATLAS Collaboration}

\bigskip

M.~Aaboud$^\textrm{\scriptsize 34d}$,    
G.~Aad$^\textrm{\scriptsize 99}$,    
B.~Abbott$^\textrm{\scriptsize 124}$,    
O.~Abdinov$^\textrm{\scriptsize 13,*}$,    
B.~Abeloos$^\textrm{\scriptsize 128}$,    
S.H.~Abidi$^\textrm{\scriptsize 165}$,    
O.S.~AbouZeid$^\textrm{\scriptsize 143}$,    
N.L.~Abraham$^\textrm{\scriptsize 153}$,    
H.~Abramowicz$^\textrm{\scriptsize 159}$,    
H.~Abreu$^\textrm{\scriptsize 158}$
% ...
\bigskip
\\

% ...
$^{13}$Institute of Physics, Azerbaijan Academy of Sciences, Baku; Azerbaijan.\\
% ...
$^{34}$$^{(a)}$Facult\'e des Sciences Ain Chock, R\'eseau Universitaire de Physique des Hautes Energies - Universit\'e Hassan II, Casablanca;$^{(b)}$Centre National de l'Energie des Sciences Techniques Nucleaires (CNESTEN), Rabat;$^{(c)}$Facult\'e des Sciences Semlalia, Universit\'e Cadi Ayyad, LPHEA-Marrakech;$^{(d)}$Facult\'e des Sciences, Universit\'e Mohamed Premier and LPTPM, Oujda;$^{(e)}$Facult\'e des sciences, Universit\'e Mohammed V, Rabat; Morocco.\\
% ...
$^{99}$CPPM, Aix-Marseille Universit\'e, CNRS/IN2P3, Marseille; France.\\
% ...
$^{124}$Homer L. Dodge Department of Physics and Astronomy, University of Oklahoma, Norman OK; United States of America.\\
% ...
$^{128}$LAL, Universit\'e Paris-Sud, CNRS/IN2P3, Universit\'e Paris-Saclay, Orsay; France.\\
% ...
$^{143}$Santa Cruz Institute for Particle Physics, University of California Santa Cruz, Santa Cruz CA; United States of America.\\
% ...
$^{153}$Department of Physics and Astronomy, University of Sussex, Brighton; United Kingdom.\\
% ...
$^{158}$Department of Physics, Technion, Israel Institute of Technology, Haifa; Israel.\\
$^{159}$Raymond and Beverly Sackler School of Physics and Astronomy, Tel Aviv University, Tel Aviv; Israel.\\
% ...
$^{165}$Department of Physics, University of Toronto, Toronto ON; Canada.\\
% ...
$^{a}$ Also at  Department of Physics, University of Malaya, Kuala Lumpur; Malaysia.\\
$^{b}$ Also at Borough of Manhattan Community College, City University of New York, NY; United States of America.\\
$^{c}$ Also at Centre for High Performance Computing, CSIR Campus, Rosebank, Cape Town; South Africa.\\
$^{d}$ Also at CERN, Geneva; Switzerland.\\
$^{e}$ Also at CPPM, Aix-Marseille Universit\'e, CNRS/IN2P3, Marseille; France.\\
% ...
$^{*}$ Deceased

\end{flushleft}


\end{document}